# Large eddy simulation of a supersonic lifted hydrogen flame with sparse-Lagrangian multiple mapping conditioning approach

Zhiwei Huang[1], Matthew J. Cleary[2], Zhuyin Ren[3], Huangwei Zhang[1*]

[1] *Department of Mechanical Engineering, National University of Singapore, 9 Engineering Drive 1, Singapore 117576, Singapore*
[2] *School of Aerospace, Mechanical and Mechatronic Engineering, The University of Sydney, New South Wales 2006, Australia*
[3] *Institute for Aero Engine, Tsinghua University, Beijing, 100084, China*

## Abstract

The Multiple Mapping Conditioning / Large Eddy Simulation (MMC-LES) approach is used to simulate a supersonic lifted hydrogen jet flame, which features shock-induced autoignition, shock-flame interaction, lifted flame stabilization, and finite-rate chemistry effects. The shocks and expansion waves, shock-reaction interactions and overall flame characteristics are accurately reproduced by the model. Predictions are compared with the detailed experimental data for the mean axial velocity, mean and root-mean-square temperature, species mole fractions, and mixture fraction at various locations. The predicted and experimentally observed flame structures are compared through scatter plots of species mole fractions and temperature against mixture fraction. Unlike most past MMC-LES which has been applied to low-Mach flames, in this supersonic flame case pressure work and viscous heating are included in the stochastic FDF equations. Analysis indicates that the pressure work plays an important role in autoignition induction and flame stabilization, whereas viscous heating is only significant in shear layers (but still negligibly small compared to the pressure work). The evolutions of particle information subject to local gas dynamics are extracted through trajectory analysis on representative fuel and oxidizer particles. The particles intermittently enter the extinction region and may be deviated from the full burning or mixing lines under the effects of shocks, expansion waves and viscous heating. The chemical explosive mode analysis performed on the Lagrangian particles shows that temperature, the H and OH radicals contribute dominantly to CEM respectively in the central fuel jet, fuel-rich and fuel-lean sides. The pronounced particle Damköhler numbers first occur in the fuel jet / coflow shear layer, enhanced at the first shock intersection point and peak around the flame stabilization point.

---

* Corresponding author. E-mail: huangwei.zhang@nus.edu.sg. Tel: +65 6516 2557.

# 1. Introduction

The fundamental physics of supersonic combustion have attracted much attention from researchers in recent years, driven by the rapid, ongoing development of hypersonic propulsion technology [1–4]. Supersonic flames feature very complex interactions between turbulence, flow discontinuity (e.g., shocks), and chemistry. However, understanding of these phenomena remains incomplete. While useful and continuously improving, experiments under conditions relevant to supersonic engines are expensive and extracting detailed measurements is difficult [3]. Computational Fluid Dynamics (CFD) simulations can and should play a vital complementary role in exploring the physics [1,2]. In particular, Large Eddy Simulation (LES) has the potential to provide detailed spatiotemporal information on supersonic combustion processes at increasingly affordable cost, and consequently it has been used for modelling both fundamental and applied configurations, e.g. coflow [5–7] and crossflow [8–10] jet flames, and model combustors [11–13]. Advanced Sub-Grid Scale (SGS) combustion models are required to accurately capture the unresolved turbulent fluctuations of reaction rates and their interactions with turbulence and characteristic structures in high-speed flows, including shocks and expansion waves [1]. Among the available SGS models, the Probability Density Function (PDF) [14] approaches are the most universal in their range of applicability, because the non-linear chemical source terms are closed naturally independent of the specific turbulence model or flame regime [14]. In LES, solutions are obtained for the so called Filtered Density Function (FDF) [15]. Application of FDF method to supersonic flows has been relatively rare, but includes the stochastic fields FDF approach of De Almeida et al. [7] and the non-reacting fundamental work of Zhang et al. [16].

The two major challenges in FDF methods lie in finding good models for the SGS micro-mixing term and in reducing their relatively high computational cost [17–20]. The Multiple Mapping Conditioning approach (MMC) [21,22] aims to tackle both of these challenges. In its stochastic form, MMC is a full transported PDF / FDF method [18] which introduces concepts from the Conditional Moment Closure (CMC) model [23] to condition the mixing model to ensure that it is local in composition space (a vital property of mixing [24]). This is done indirectly though localizing in a mathematically independent reference space so that two other vital mixing properties (i.e. independence and linearity for all scalars [24]) are preserved. In LES of non-premixed flames, the filtered mixture fraction solved in Eulerian fashion on the LES mesh is an appropriate reference variable since it effectively parameterizes the composition while also being mathematically independent of the composition field on the particles [22]. The enforced localness in Reference Mixture Fraction (RMF) space permits a relaxation of strict mixing localness in physical space and a reduction in the number of Lagrangian particles compared to approaches with conventional (non-local) mixing models.

Sparse-Lagrangian MMC-LES involves significantly fewer particles for the stochastic composition field than Eulerian cells for the LES solution [19]. Therefore, the computational cost is significantly lower than conventional FDF methods with an intensive distribution of particles. Sparse MMC-LES has been extensively validated with a range of subsonic experimental combustion configurations of practical relevance, e.g. piloted methane / air jet diffusion flames [25], methane / air swirl flames [26], Sandia DME flame series [27] and turbulent spray flames [28]. In our recent work, sparse MMC-LES was extended to highly-compressible conditions for the first time [29] and demonstrated good accuracy against experimental data [30] for velocity, pressure and temperature data in a model supersonic combustor. The roles of pressure work and viscous heating in flame stabilization were preliminarily analyzed and it was found that the pressure work played a significant role in the unsteady behavior of the flame base whereas the viscous heating was negligibly small. However, due to the limited availability of data from the experimental combustor, validation of MMC-LES predictions of reactive scalars was not possible.

The objective of the present work is to model an autoignition stabilized supersonic hydrogen jet flame for which detailed experimental data of reactive scalars is available [31]. This very well-characterized target flame features a broad range of physical phenomena, including shock-induced autoignition, shock-flame interaction, lifted flame stabilization, and finite-rate chemistry effects. Therefore, it will be helpful for more comprehensive examination of the compressible MMC-LES model. Moreover, a number of other novel contributions will be made here. Firstly, the importance of pressure work and viscous heating on autoignition and flame stabilization in MMC are re-assessed in detail and the aforementioned negligibility of the latter is questioned. Secondly, the Lagrangian evolutions of temperature and mixture fraction are extracted to study the interactions between unsteady flame behaviors and hypersonic gas dynamics. Thirdly, Chemical Explosive Mode Analysis (CEMA) [32] is used to extract the accurate chemical information, including chemical reaction timescale and Damköhler number, from different flame sections.

The rest of the paper is organized as follows. The formulation and implementation of the compressible MMC-LES are detailed in Section 2. The experimental and numerical configurations are presented in Sections 3 and 4, respectively. The results and discussion are given in Section 5, followed by the conclusions in Section 6.

## 2. Governing equation and numerical implementation

### 2.1. Compressible Eulerian LES equation

The filtered equations for mass, momentum and reference mixture fraction are solved using Eulerian LES. The filtered continuity equation is

$$\frac{\partial \bar{\rho}}{\partial t} + \frac{\partial}{\partial x_j}\left(\bar{\rho}\tilde{u}_j\right) = 0, \tag{1}$$

where $t$ is time, $x_j$ is the spatial coordinate, $\bar{\rho}$ is filtered density and $\tilde{u}_j$ is Favre filtered $j$-th velocity component. The filtered momentum equation is

$$\frac{\partial}{\partial t}(\bar{\rho}\tilde{u}_i) + \frac{\partial}{\partial x_j}(\bar{\rho}\tilde{u}_i\tilde{u}_j) + \delta_{ij}\frac{\partial \bar{p}}{\partial x_j} - \frac{\partial}{\partial x_j}\left(\tilde{\tau}_{ij} - \tau_{ij}^{sgs}\right) = 0, \tag{2}$$

where $\bar{p}$ is filtered pressure, $\delta_{ij}$ is a Kronecker delta function and $\tilde{\tau}_{ij}$ is the molecular viscous stress tensor, i.e.,

$$\tilde{\tau}_{ij} = \mu\left(\frac{\partial \tilde{u}_i}{\partial x_j} + \frac{\partial \tilde{u}_j}{\partial x_i} - \frac{1}{3}\delta_{ij}\frac{\partial \tilde{u}_k}{\partial x_k}\right). \tag{3}$$

Here $\mu$ is the dynamic viscosity, which is modelled by Sutherland's law. The SGS viscous stress tensor $\tau_{ij}^{sgs}$ in Eq. (2) is given by

$$\tau_{ij}^{sgs} = -\mu_t \left( \frac{\partial \tilde{u}_i}{\partial x_j} + \frac{\partial \tilde{u}_j}{\partial x_i} - \frac{1}{3}\delta_{ij}\frac{\partial \tilde{u}_k}{\partial x_k} \right) + \frac{1}{3}\delta_{ij}\bar{\rho}k_t, \tag{4}$$

where $k_t$ and $\mu_t$ are respectively the SGS kinetic energy and dynamic viscosity. In the present study, they are closed using the standard Smagorinsky model [33].

The filtered RMF ($\tilde{f}$) is used for localizing stochastic particle mixing. RMF is a conserved normalized scalar with $\tilde{f} = 1$ in the fuel stream and $\tilde{f} = 0$ in the oxidizer stream. Its equation reads

$$\frac{\partial}{\partial t}(\bar{\rho}\tilde{f}) + \frac{\partial}{\partial x_j}(\bar{\rho}\tilde{f}\tilde{u}_j) - \frac{\partial}{\partial x_j}\left(\bar{\rho}\mathcal{D}_{eff}\frac{\partial \tilde{f}}{\partial x_j}\right) = 0, \tag{5}$$

where $\mathcal{D}_{eff} = \mathcal{D}_m + \mathcal{D}_t$ is the sum of molecular and SGS diffusivities. The molecular diffusivity is modelled as $\mathcal{D}_m = \mu/\bar{\rho}Sc$ with Schmidt number $Sc = 0.7$ and the SGS diffusivity is $\mathcal{D}_t = \mu_t/\bar{\rho}Sc_t$ with turbulent Schmidt number $Sc_t = 0.4$ [34].

Species and standardised enthalpy are FDF state space variables solved on Lagrangian particles as discussed in Section 2.2. Mass and energy consistency between the particle fields and the Eulerian LES is achieved through solution of additional Eulerian filtered equations for the *equivalent species* and *equivalent total enthalpy* [29,35], governed by

$$\frac{\partial}{\partial t}(\bar{\rho}\tilde{Y}_m^E) + \frac{\partial}{\partial x_j}(\bar{\rho}\tilde{Y}_m^E\tilde{u}_j) - \frac{\partial}{\partial x_j}\left(\bar{\rho}\mathcal{D}_{eff}\frac{\partial}{\partial x_j}\tilde{Y}_m^E\right) = \frac{\bar{\rho}(\widetilde{Y_m|f}^E - \tilde{Y}_m^E)}{\tau_{rel}}. \tag{6}$$

and

$$\frac{\partial}{\partial t}(\bar{\rho}\tilde{H}_t^E) + \frac{\partial}{\partial x_j}(\bar{\rho}\tilde{H}_t^E\tilde{u}_j) - \frac{\partial \bar{p}}{\partial t} - \frac{\partial}{\partial x_j}\left(\bar{\rho}\mathcal{D}_{eff}\frac{\partial}{\partial x_j}\tilde{h}_s^E\right) - \frac{\partial}{\partial x_j}(\tilde{\tau}_{ij}\tilde{u}_i) = \frac{\bar{\rho}(\widetilde{h_s|f}^E - \tilde{h}_s^E)}{\tau_{rel}}, \tag{7}$$

respectively. Generally, it is computationally efficient to solve for a limited set of equivalent species which make up a significant fraction of the mixture (say 99% by mass) and thus control its thermodynamic state. The filtered total enthalpy is defined as $\tilde{H}_t^E = \tilde{h}_s^E + \frac{1}{2}\sum_{j=1}^{3}\tilde{u}_j^2$ where $\tilde{h}_s^E$ is the filtered equivalent sensible enthalpy. The third term on the LHS of Eq. (7) is the pressure work and the last term on the LHS is the viscous heating. The source terms on RHS of Eqs. (6) and (7) relax the Eulerian equivalent fields over the timescale $\tau_{rel}$ towards the conditional means in RMF space, $\widetilde{Y_m|f}^E$ and $\widetilde{h_s|f}^E$. Note that the source terms here are different from those in so called quasi-laminar LES closures of the filtered source term, which neglect SGS fluctuations of the species fields. Since the conditional means, $\widetilde{Y_m|f^E}$ and $\widetilde{h_s|f^E}$, are estimated from the stochastic particle fields (see Section 2.3) the equivalent species and enthalpy source terms therefore explicitly include SGS fluctuations which, in non-premixed combustion, are largely driven by the fluctuations in the mixture fraction.

Finally, the Eulerian pressure is obtained through the ideal gas equation of state,

$$\bar{p} = \bar{\rho} R_u \tilde{T}^E \sum_{m=1}^{N_E} \frac{\tilde{Y}_m^E}{M_m}, \tag{8}$$

where $R_u$ = 8.314 J/(mol·K) is the universal gas constant, $M_m$ and $\tilde{Y}_m^E$ are, respectively, the molecular weight and filtered mass fraction of $m$-th equivalent species, $N_E$ is the total number of equivalent species that are solved, and $\tilde{T}^E$ is the equivalent temperature updated from the equivalent enthalpy.

## 2.2. Compressible stochastic differential equations on Lagrangian particles

The equivalent set of Stochastic Differential Equations (SDEs) for the evolution of the joint FDF of species mass fractions and standardised enthalpy are solved on an ensemble of Lagrangian notional particles [21,29]. They read

$$dx_i^q = \left[\tilde{u}_i + \frac{1}{\bar{\rho}} \frac{\partial}{\partial x_i}(\bar{\rho} \mathcal{D}_{eff})\right]^q dt + \delta_{ij}\left(\sqrt{2\mathcal{D}_{eff}}\right)^q d\omega_i, \tag{9}$$

$$dY_m^q = \left(W_m^q + S_m^q\right)dt, \tag{10}$$

$$dh^q = \left[W_h^q + S_h^q + \left(\frac{1}{\bar{\rho}} \frac{D\bar{p}}{Dt}\right)^q + \left(\frac{1}{\bar{\rho}} \tilde{\tau}_{ij} \frac{\partial \tilde{u}_i}{\partial x_j}\right)^q\right] dt, \tag{11}$$

$$\langle S^{p,q}|\tilde{f}, \boldsymbol{x}\rangle = 0. \tag{12}$$

Here $q$ is a particle index associated with a stochastic realization of the turbulent field. Equation (9) is for transport of particles in physical space where $x_i$ is the $i$-th component of the position vector and $d\omega_i$ is the increment of an independent Wiener process. Equations (10) and (11) govern the evolution of particle mass fractions, $Y_m^q$, and standardised enthalpy, $h^q = \left(h_f^\theta + \int_{T_0}^T C_p dT\right)^q$, respectively, with $h_f^\theta$ being the enthalpy of formation. $W_m^q$ is the closed non-linear chemical source term and $W_h^q$ is radiative heat loss term (set to zero here). $S_m^q$ and $S_h^q$ are the mixing terms for the dissipation of conditional subfilter fluctuations of mass fraction and standardised enthalpy, respectively, and Eq. (12)

represents the MMC model constraint that conserves conditional means during mixing through enforced localness in the combined $(\pmb{x}, \tilde{f})$-space. The particular form of the mixing operation used here adopts a variant of the Curls mixing model [36]. Particles are mixed in pairs (particles *p* and *q*), where the mean distance between the mixing pairs in $(\pmb{x}, \tilde{f})$-space is controlled by global parameters $r_m$ (the characteristic distance in $\pmb{x}$-space) and $f_m$ (the characteristic distance in $\tilde{f}$-space). Here $f_m = 0.01$ [37] and $r_m$ is obtained by the fractal model [22]. The pairwise mixing is linear and has a mixing time scale, $\tau_L$, which is modelled by the a-ISO model [38]. Different from the previous MMC model for low-Mach flows [21,35], the pressure work, $W_P^q = \left(\frac{1}{\bar{\rho}}\frac{D\bar{p}}{Dt}\right)^q$, and viscous heating, $W_{VH}^q = \left(\frac{1}{\bar{\rho}}\tilde{\tau}_{ij}\frac{\partial \tilde{u}_i}{\partial x_j}\right)^q$, are incorporated in Eq. (11) and their roles in predicting supersonic flames are preliminarily discussed in our recent work [29].

*2.3. Numerical implementation*

The numerical implementation of the hybrid Eulerian/Lagrangian compressible MMC-LES model into the *mmcFoam* suite of solvers [35] was recently presented in [29]. The Eulerian scheme is based on the *RYrhoCentralFoam* solver [39–41] which has been validated previously for a range of benchmark problems, including Sod's shock tube, a forward-facing step, a supersonic jet and shock-vortex interaction [39,40]. The stochastic Lagrangian particle implementation is presented and extensively validated in [35] and has been augmented to include compressible pressure work and viscous heating terms in [29]. Essential details of the schemes are given below.

The finite volume forms of the Eulerian LES equations for momentum, RMF, equivalent species mass fraction, and equivalent total enthalpy (i.e., Eqs. 2, 5, 6, 7, respectively) are integrated with an operator-splitting method [29,39]. A second-order implicit Crank-Nicolson scheme is used for discretizing the unsteady terms and the MUSCL-type KNP scheme [42] with Minmod flux limiter [43] is used to discretize the convective terms. The diffusive fluxes are predicted with a second-order central differencing scheme.

The stochastic Lagrangian transport equations (Eqs. 9-12) are integrated as three fractional steps. Spatial transport in Eq. (9) uses the first-order Euler-Maruyama scheme [44]. Chemical source terms in Eqs. (10)-(11) are integrated using a stiff ODE solver *seulex* [45]. The particle pairs for mixing in Eq. (12) are selected dynamically using a k-dimensional tree algorithm [46].

The Eulerian and Lagrangian parts of the *mmcFoam* solvers are two-way coupled. In the *forward coupling* step, in order to integrate Eqs. (9) and (11) with the constraint (12), the filtered velocity, pressure material derivative, viscous heating, RMF, effective (molecular plus SGS) diffusivity and its gradient are all tri-linearly interpolated from the underlying Eulerian LES fields to the particle locations. In the *backward coupling* step, mass and energy consistency between the two schemes is achieved by passing estimates of $\widetilde{Y_m|f^E}$ and $\widetilde{h_s|f^E}$ from the stochastic particle fields to the Eulerian scheme to solve Eqs. (6) and (7). This estimation involves integration over stochastic particles around each Eulerian grid with weighting by radial basis functions in both RMF and physical space. Full details are provided in [35].

## 3. Experimental configuration

Figure 1(a) shows the schematic of the supersonic hydrogen jet flame experimentally investigated by Cheng et al. [31]. The burner consists of a central round fuel pipe with diameter $D_f = 2.36$ mm issuing a sonic hydrogen jet surrounded by an annular hot vitiated coflow at Mach 2.0 generated by an upstream lean-burned hydrogen combustor. Figure 1(b) shows the details of the fuel jet and coflow nozzle near the burner exit. The resulting flame is stabilized at about $25D_f$ downstream of the nozzle exit (red line in Fig. 1c), inferred from the experimental visual photograph of flame luminosity in Fig. 1(c). The major dimensions of the burner and the flow conditions at the exits of both hydrogen and coflow streams are detailed in Table 1. Small concentrations of radical species, such as OH, are also present in the vitiated coflow [31], but are neglected in the modelling and only the major product species fractions are specified at the inflow boundaries.

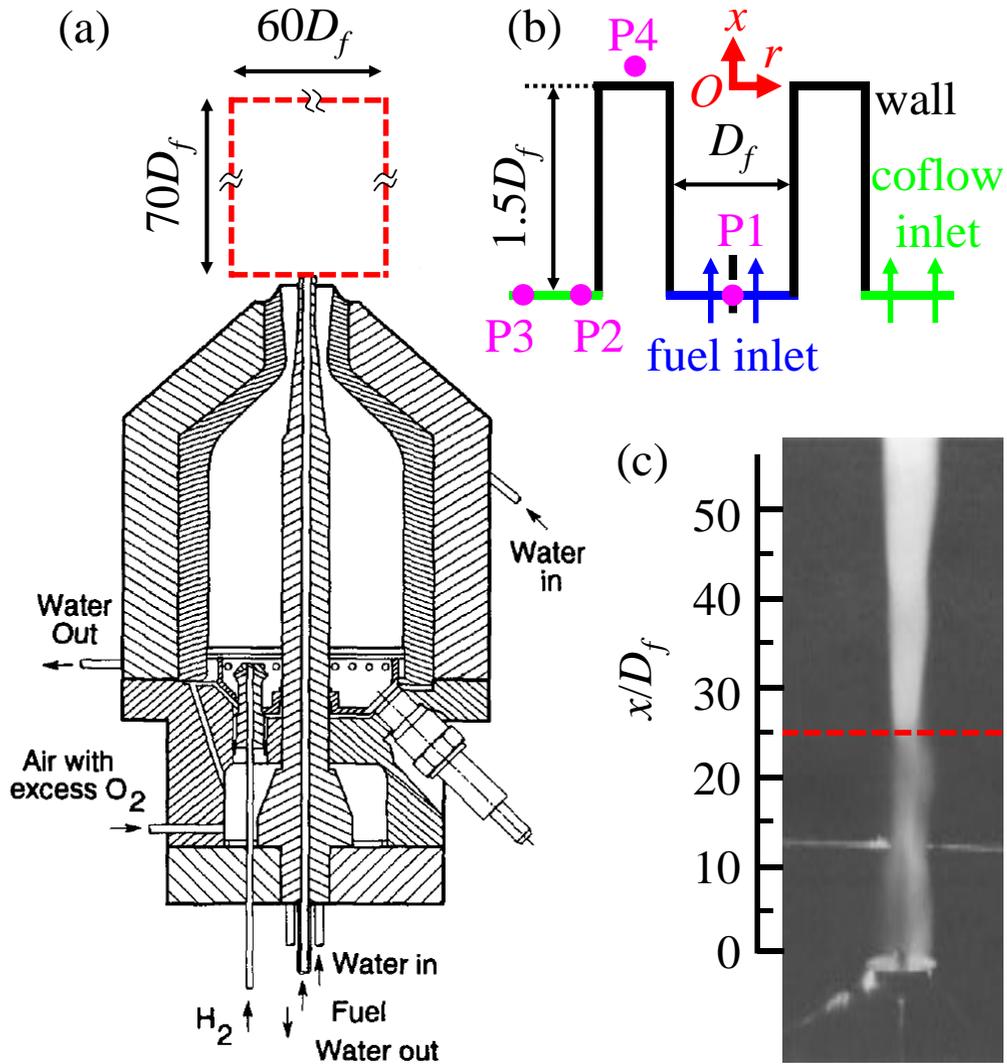

**Fig. 1.** (a) Schematic of the experimental configuration [31], (b) detail around the burner exit, and (c) long exposure visual photograph [31].

**Table 1.** Major dimensions and boundary conditions for the Cheng supersonic flame [31]. $D_i$ is inner diameter, $D_o$ is outer diameter, $p$ is pressure, $T$ is temperature, $Re$ is Reynolds number, $Ma$ is flow Mach number, $X_m$ is mole fraction of the $m$-th species.

| | $D_i$ [mm] | $D_o$ [mm] | $p$ [kPa] | $T$ [K] | $Re$ | $Ma$ | $X_{O2}$ | $X_{H2O}$ | $X_{N2}$ | $X_{H2}$ |
|---|---|---|---|---|---|---|---|---|---|---|
| Hydrogen | 0 | 2.36 | 112 | 545 | 15,600 | 1.0 | 0.0 | 0.0 | 0.0 | 1.0 |
| Coflow | 3.81 | 17.78 | 107 | 1,250 | 101,100 | 2.0 | 0.201 | 0.255 | 0.544 | 0.0 |

## 4. Numerical configuration

*4.1. Eulerian computational configuration*

As shown in Fig. 1(a), the cylindrical computational domain is $70D_f$ in the streamwise direction ($x$) by $30D_f$ in radial direction ($r$). The coordinate origin lies at the center of the fuel jet exit (i.e., point $O$ in Fig. 1b). The mesh is discretized by 12,175,200 hexahedron cells with refinement around the fuel jet, fuel / coflow shear layer and across the coflow with a minimum cell size of 118 μm in the $x$-direction by 50 μm in $r$-direction, respectively. As a comparison, the Kolmogorov length scale for this flow is estimated to be the 10-20 μm range and the integral length scale is of order 3-7 mm [31]. Therefore, the finest cell in the present work is between approximately three and seven times the Kolmogorov length scale. *A posterior* analysis of the LES mesh resolution (see Section A of supplementary document) shows that the present LES resolution is sufficient for predicting the kinetic energy (more than 90% is resolved) and scalar variations in the jet flame. Moreover, the maximum CFL (Courant-Friedrichs-Lewy) number is 0.1, which approximately corresponds to the physical time step of $10^{-9}$ s.

The upstream boundary of the cylindrical computational domain (see Fig. 1a) extends $x/D_f = -1.5$ (i.e. 3.54 mm) into the fuel and coflow inlet pipes (Fig. 1b) where Dirichlet conditions are enforced and, in line with experimental observations, turbulent velocity fluctuations of about 18% and 22% of the mean values are applied through a synthetic turbulence generator [47] and the Reynolds stress is given following the method of Masri et al. [48] and Zhang et al. [49]. Based on the work of Bouheraoua et al. [6] and Zhao et al. [50] on the same flame, the inflow turbulence is significant for the shear layer development after the nozzle exits and near-field shock structures. An adiabatic no-slip wall condition is adopted at the fuel and coflow injector lips. The spatially averaged $y^+$ values are 0.67 and 1.64 near the fuel and coflow pipe walls, respectively. Therefore, the near-wall turbulence is well resolved. Similar wall treatment has been adopted in Ref. [6], and satisfactory results are obtained with a comparable mesh resolution to ours. Non-reflective conditions [51] are applied at the circumferential boundary of the domain. Since the outflow is supersonic, zero gradient conditions are applied for all variables.

*4.2. Lagrangian computational configuration*

Nominally 760,000 Lagrangian particles are used in the domain, corresponding to a sparse distribution of about one Lagrangian particle for every 16 Eulerian LES cells (1L/16E). The resultant characteristic spacing between particles in the central jet and shear layer is approximately $\Delta_L = 0.12$ mm. The low sensitivity of MMC-LES to increased resolution has been extensively demonstrated for subsonic jet flames [26,35], and more recently for supersonic combustion as well [29]. The particle resolution used in the present work is comparable to the finest case in our recent study with $\Delta_L = 0.115$ mm [29].

A detailed hydrogen mechanism, containing 9 species ($H_2$, $O_2$, $N_2$, $H_2O$, $HO_2$, $H_2O_2$, H, O and OH) and 19 elementary reactions [52], is used. Validations have shown that it can well reproduce the measured ignition delay and laminar flame speed over a range of pressure conditions [53].

In the Eulerian equivalent composition equations (6)-(7) and stochastic particle equations (10)-(11), unity Lewis number is used for all the species, whereas the molecular and turbulent Prandtl numbers are 0.71 and 0.9, respectively [5–7,51,54].

## 5. Results and discussion

Following a statistically transient flow period of 0.3 ms to eliminate the effects of the initial conditions, the stationary statistics presented in this section were integrated over 0.7 ms corresponding to about six characteristic flow-through times given by the ratio of the domain length ($70D_f$) and co-flow bulk velocity (1,420 m/s). In the following, instantaneous filtered quantities are represented by tilde (e.g., $\tilde{f}$) and the time-averaged quantities are represented by angle brackets (e.g., $\langle f \rangle$).

*5.1. Flow and flame structures*

Figure 2 shows various aspects of the predicted flow and flame structures. Contours of instantaneous and mean pressure gradient magnitudes are shown in Figs. 2(a) and 2(b), respectively. Iso-lines of instantaneous stoichiometric reference mixture fraction, $\tilde{f}_{st} = 0.0297$, are also shown in Fig. 2(a). The shock structures immediately downstream of the nozzle exit show a good resemblance to those captured respectively by Moule et al. [5] whose simulations extended upstream and included detailed nozzle geometry, Bouheraoua et al. [6] who used synthetic inflow turbulent velocity boundary conditions at the nozzle exit plane, and Zhao et al. [50] who made detailed comparison with and without synthetic inflow velocity fluctuations. In comparison with the results in Refs. [40,50] produced without a turbulent inflow, the Mach trains shown in Fig. 2(a) are less intense (lower contrast) and become more diffuse (blurred) in the far field. This is because the shock-turbulence interactions and more rapid turbulent jet breakup resulting from the realistic inflow turbulent spectrum modulate the development of very strong shock wave structures. The diamond-like shock structures are clearly identified in Fig. 2(b) through the mean pressure gradient. Expansion waves (E1) form at the fuel jet exit with radially adjacent shocks (S1). These are reflected from the shear layer between the vitiated coflow and the surrounding ambient air. Their interaction forms a Mach disk after the burner exit and leads to the distinctive diamond-shaped alternating expansion / shock wave system (E1-E3 and S1-S3). The shock strength decays with downstream distance due to turbulence-shock-flame interactions. This is most obvious downstream of $x/D_f = 25$ (line $l_{fb}$), where the flame base is stabilized near the intersection of two shocks indicated by point 'b'. It is possible that the onset of substantial heat release and hence thermal expansion weakens the shock waves substantially [55,56].

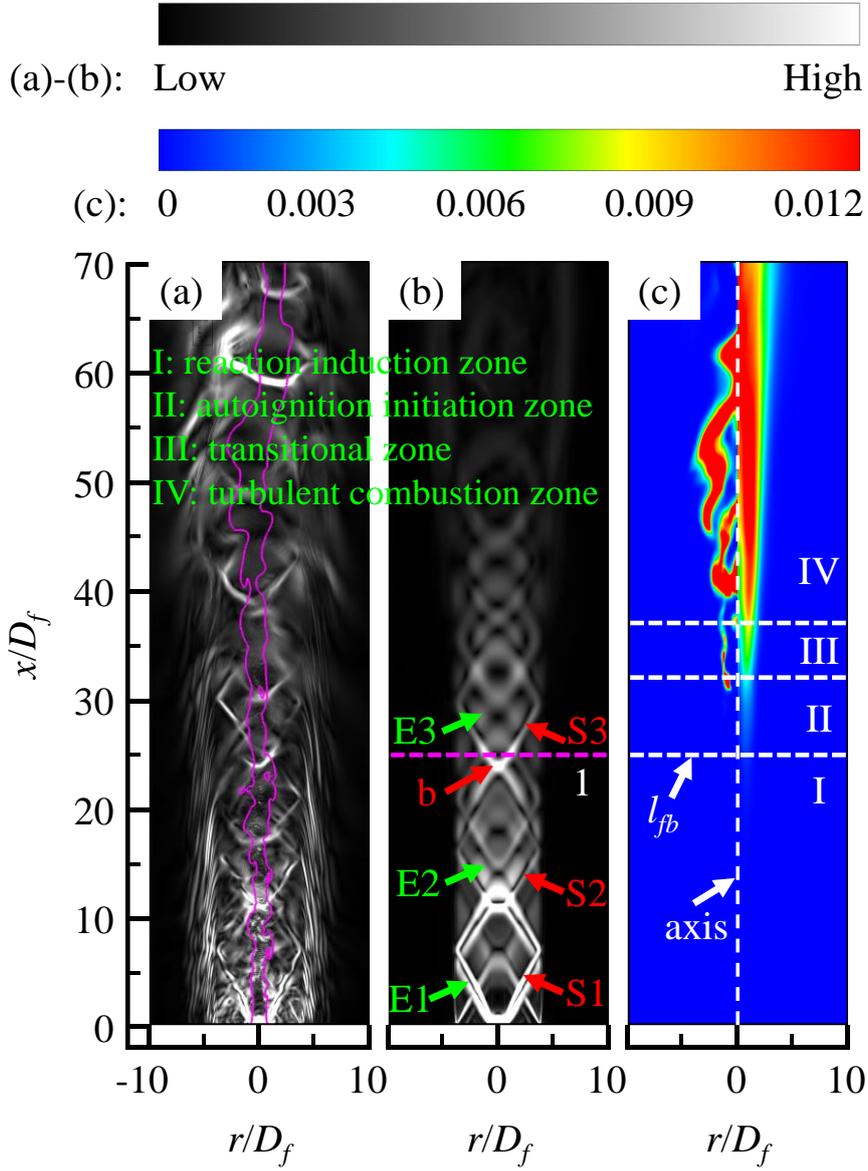

**Fig. 2.** Flow and flame structures: (a) instantaneous pressure gradient magnitude, (b) mean pressure gradient magnitude and (c) instantaneous (left) and mean (right) equivalent OH mole fraction. Iso-lines in (a) represent instantaneous stoichiometric mixture fraction.

Figure 2(c) shows the instantaneous equivalent OH mole fraction fields $\tilde{X}_{OH}^{E}$, and its mean $\langle X_{OH}^{E} \rangle$, on the left and right sides of the image, respectively. Note that the equivalent species fields are useful for the purpose of visualization whereas the stochastic particle fields are the real quantities used below for direct quantitative comparison against the experiment. The mean lift-off distance corresponding to $\langle X_{OH}^{E} \rangle = 0.008$ is at about $x/D_f = 25$ (line $l_{fb}$), which agrees well with the experimental value (c.f. Fig. 1c). Upstream of line $l_{fb}$, there is no observable OH radical, but extensive HO$_2$ is generated (see Fig.

3), which is a precursor species of autoignition. Hence, this region is regarded as the *reaction induction zone* (Region I in Fig. 2c). For $25 < x/D_f < 32$, $\langle X_{OH}^E \rangle$ is still rather small and this region is called the *autoignition initiation zone* (Region II). For $32 < x/D_f < 37$, $\langle X_{OH}^E \rangle$ grows rapidly in the *transitional zone* (Region III), and downstream of $x/D_f = 37$ in the *turbulent combustion zone* (Region IV) $\langle X_{OH}^E \rangle$ is much larger still. The staged flame development shows close qualitative similarity to that experimentally observed (again, c.f. Fig. 1c). This will be further analyzed in Section 5.4.

Figure 3 shows instantaneous $HO_2$ and OH mole fractions as well as mixture fraction and temperature on the Lagrangian particles. Note that zones I-IV are the same as those in Fig. 2. As indicated by circle 'A' in Fig. 3(a), isolated $HO_2$ pockets are first observed in the shear layers between the fuel jet and coflow in the reaction induction zone (Region I) and begins to accumulate into a continuous field downstream of point 'a' where the shocks first intersect along the jet centerline. OH radical in Fig. 3(b) appears in very isolated pockets indicated by circles 'B' and 'C' near the flame base (line $l_{fb}$) well downstream of where $HO_2$ first appears, and peaks in turbulent combustion zone (Region IV). Its spatial distribution reflects the fact that OH is produced in the thin stoichiometric reaction layers and consumed by slow recombination reactions, which is significant in the main reaction zones (i.e., the transitional and turbulent combustion zones, Regions III and IV). This is also demonstrated in the mean Eulerian fields in Fig. 2(c). For the instantaneous mixture fraction in Fig. 3(c), as is to be expected, the breakup of the turbulent fuel jet occurs further upstream than it does in the simulations reported in [40] without inflow turbulence, and the mixing between the jet and coflow is also much enhanced. The staged distributions of *T* in Fig. 3(d) are qualitatively similar to that of $X_{OH}$.

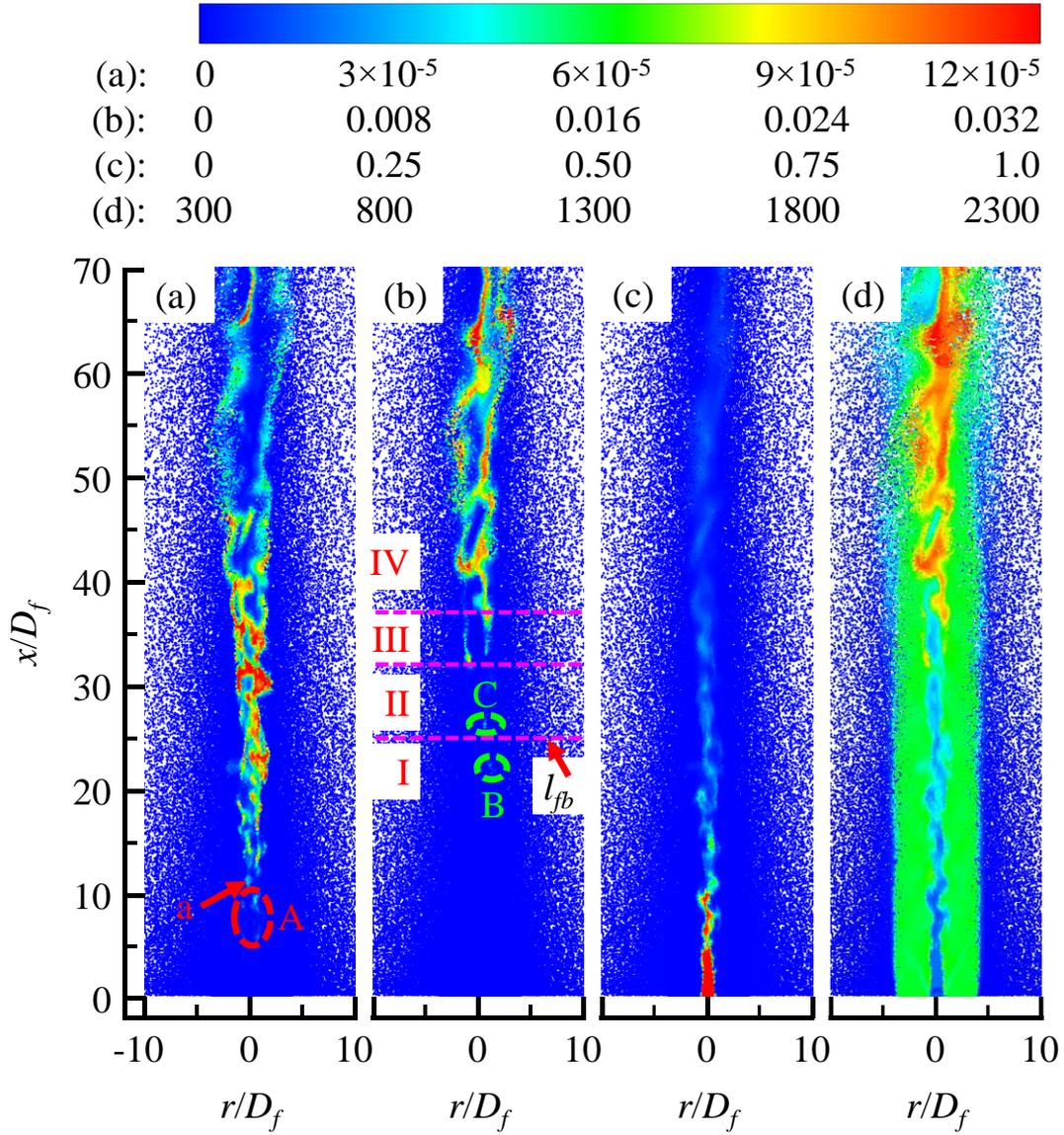

**Fig. 3.** Lagrangian particles colored by: (a) $X_{HO2}$, (b) $X_{OH}$, (c) $z$ and (d) $T$ (in K).

## 5.2. Statistics of velocity and scalars in physical space

Figure 4 shows the radial profiles of mean axial velocity, $\langle u_x \rangle$, at four streamwise locations. The overall agreement of the predictions with the experimental data is good, although there are some discrepancies. There is slight over-prediction at $x/D_f = 10.8$ near the center of the jet where the experimental data exhibits local minima. This may be caused by differences between the model and experimental turbulent velocity boundary conditions at the jet inlet. Although synthetic inflow velocity fluctuations [47] are used in the present simulations, the degree to which they mimic the experimental inflow velocity profiles cannot be quantified due to lack of data. It is noted that similar over-prediction at this location has also been observed in other studies either with [6,50] or without [40] inflow turbulence. Another discrepancy in Fig. 4 is that the model produces slightly less radial spreading of the coflow into the ambient air. This could be the result of an insufficient mesh resolution in the outer shear layer, especially at downstream locations. However, these relatively small differences in the velocity profiles are expected to have limited influence on the predictions of autoignition and lifted flame stabilization along the centerline of the jet.

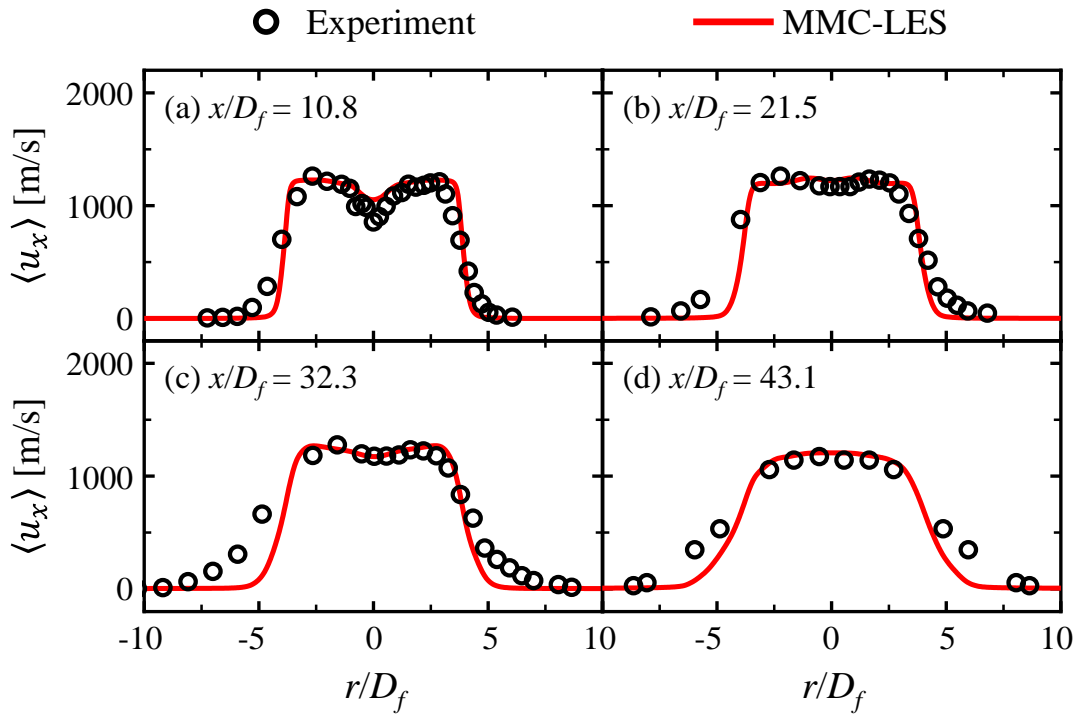

**Fig. 4.** Radial profiles of mean axial velocity. Experimental data from Ref. [31].

Figure 5 shows the radial profiles of mean temperature, $\langle T \rangle$, at various streamwise locations. Note that the experimental profiles (also for the species mole fractions in Figs. 8 and 9) are asymmetric due to the imperfect orientation of the burner [31]. Considering this uncertainty, $\langle T \rangle$ is reasonably well reproduced with MMC-LES. At $x/D_f = 21.5$, $\langle T \rangle$ is under-predicted (about 40%). Statistically, no flame (but random and isolated $HO_2$ pockets) occurs before $x/D_f = 25$ in the simulation as seen in Figs. 2-3. Therefore, $\langle T \rangle$ in the central jet is under-predicted. The experimental data shows a temperature of above 1,500 K in the central jet, which indicates early ignition before $x/D_f = 25$. Since our mean lift-off height (based on OH criterion) is fairly close to that of the measured one (i.e., $25D_f$), this under-prediction is probably because some highly transient and localized autoignition events at this particular location that are not well captured. It may be associated with two factors: (1) the unaligned fuel / coflow axes (hence unexpected turbulence and then early ignition) and (2) radicals in the coflow (from the pre-burned hydrogen-lean combustor for preheating of the coflow) [31]. For the latter, there is about 0.1% mole fraction of OH at $x/D_f = 0.85$ in the experiment, which may reduce the ignition delay time of the mixture downstream [31]. However, these two experimental uncertainties are not quantified in the experiments and hence difficult to be reproduced in our simulations. The sensitivity of the ignition delay time to OH concentration is analyzed using zero-dimensional perfectly stirred reactor model (see Section B of supplementary document). It is found that when small amount of OH (e.g., $5\times10^{-4}$ by volume, close to the experimental values in [31]) is added, the ignition delay is characterized by an abrupt decrease at a specific range of mixture temperature. As pressure is elevated (e.g., by shock compression in our flame), the sudden change of ignition delay occurs at a higher temperature range. This result, to some degree, confirms the strong dependence of the flame ignition behaviors to high-enthalpy flow conditions. Further downstream like $x/D_f = 64.7$, $\langle T \rangle$ agrees well with the experimental data except that the boundary of high temperature zone is slightly narrower. Insufficient radial spreading has been observed with various combustion models [5–7,40,51,54], which may be caused by the insufficient mesh resolution near the coflow shear layers.

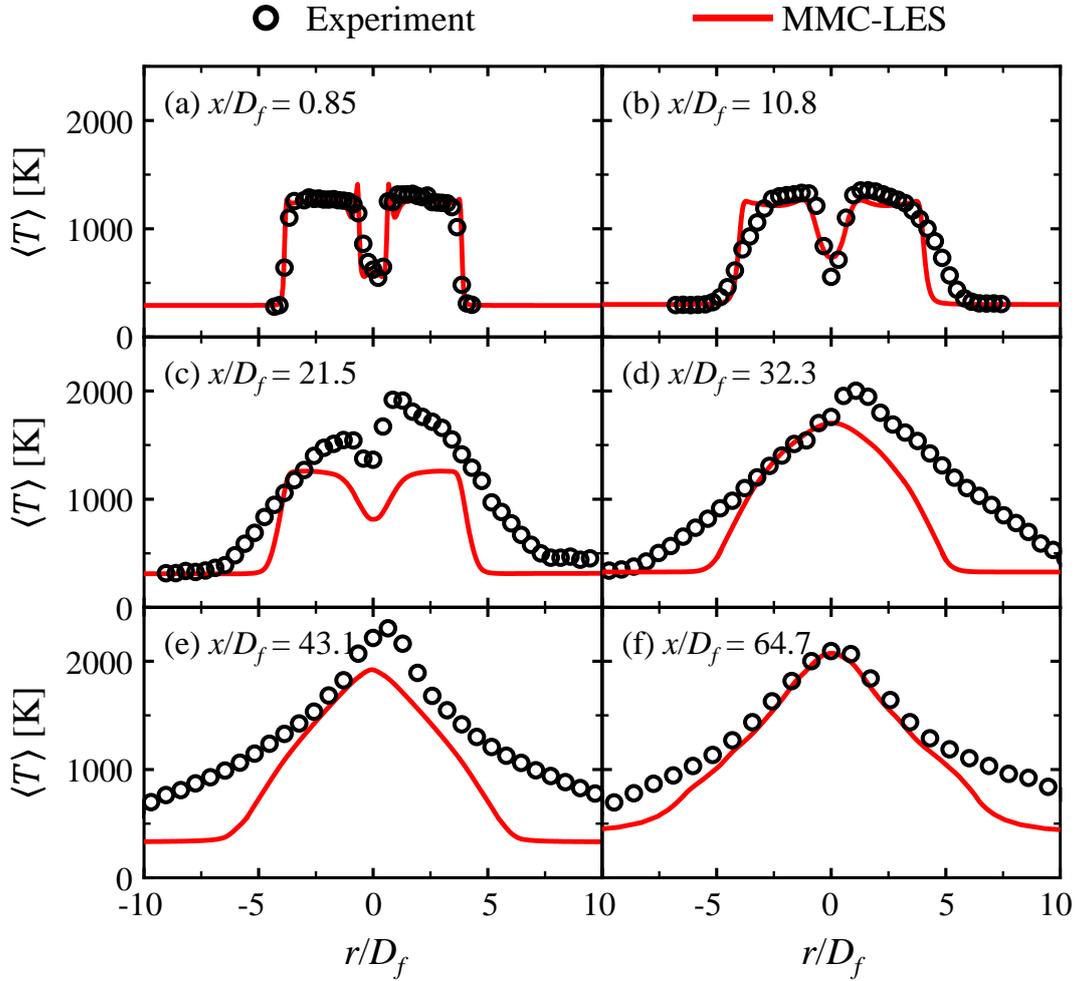

**Fig. 5.** Radial profiles of mean temperature. Experimental data from Ref. [31].

Figure 6 shows the radial profiles of temperature RMS, $\langle T^{rms} \rangle$, at the same locations. Overall, $\langle T^{rms} \rangle$ in the jet and shear layers are reasonably reproduced at $x/D_f$ = 0.85, 10.8, 43.1 and 64.7, although it is under-predicted inside the coflow at $x/D_f$ = 0.85 and 10.8 in MMC-LES. These discrepancies may be caused by the temperature fluctuation by local turbulence or chemical reactions in the radical-containing coflow in the experiment [31], as discussed in Fig. 5. Again, one can see that it is of great significance to quantify the conditions (e.g., temperature or chemical composition fluctuations) of high enthalpy flows in supersonic combustion experiments, although it is difficult in the experiments [3,31]. At $x/D_f$ = 21.5 (in the reaction induction zone I in Fig. 2), no obvious combustion occurs in our simulation. Hence, in the shear layer between the central jet and coflow, $\langle T^{rms} \rangle$ is relatively weak. At $x/D_f$ = 32.3 (in zone III), significant combustion occurs and the local $\langle T^{rms} \rangle$ increases and is quantitatively close to the measured data. Improvements are also seen for the further downstream locations in Figs. 6(e) and 6(f). However, consistent with the mean values discussed above, $\langle T^{rms} \rangle$ near the coflow shear layers (e.g., $|r|/D_f > 4$) is generally under-predicted.

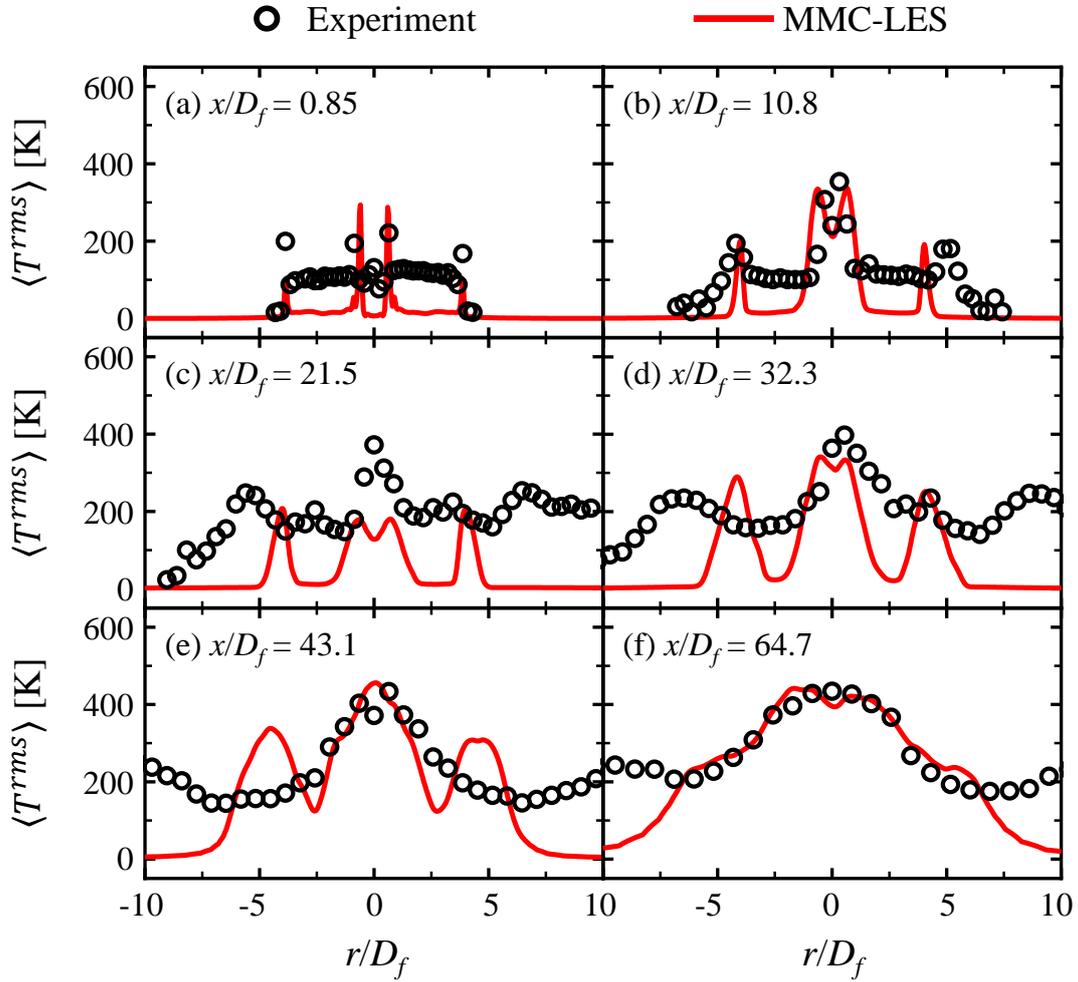

**Fig. 6.** Radial profiles of temperature RMS. Experimental data from Ref. [31].

Figures 7-9 show the radial profiles of mean species mole fractions, $\langle X_{H_2} \rangle$, $\langle X_{O_2} \rangle$, $\langle X_{H_2O} \rangle$, $\langle X_{N_2} \rangle$ and $\langle X_{OH} \rangle$, and mixture fraction, $\langle z \rangle$, at $x/D_f$ = 10.8, 21.5 and 32.3, respectively. At $x/D_f$ = 10.8 in Fig. 7, the results agree well with the experimental data, except some over-predictions of $\langle z \rangle$ in the central jet. At $x/D_f$ = 21.5 in Fig. 8, $\langle X_{OH} \rangle$ is nearly zero in our simulation, and accordingly there is little $H_2$ consumption in the central jet. This leads to the slight over-prediction of $\langle X_{H_2} \rangle$ in Fig. 8(a), but overshoot of $\langle X_{H_2O} \rangle$ in Fig. 8(c). At $x/D_f$ = 32.3 in Fig. 9, the over-prediction of $\langle X_{H_2} \rangle$ and under-prediction of $\langle X_{H_2O} \rangle$ in the central jet becomes more obvious, but the results from MMC-LES are still

of comparable accuracy with other studies, e.g., in Refs. [5,6,40,54]. This discrepancy is associated with the under-prediction of $\langle T \rangle$ at this location as shown in Fig. 5(d). Lower temperature generally indicates weaker combustion in the simulation, and therefore slower $H_2$ consumption rate and lower $H_2O$ production rate. More scalar comparisons (including mean species mole fractions and mixture fraction at $x/D_f$ = 0.85, 43.1 and 64.7, as well as RMS species mole fractions and mixture fraction at $x/D_f$ = 21.5, 32.3 and 64.7) with the experimental data are provided in Section C of supplementary document. Overall, the MMC-LES simulation well reproduces the statistics of reactive scalars in physical space.

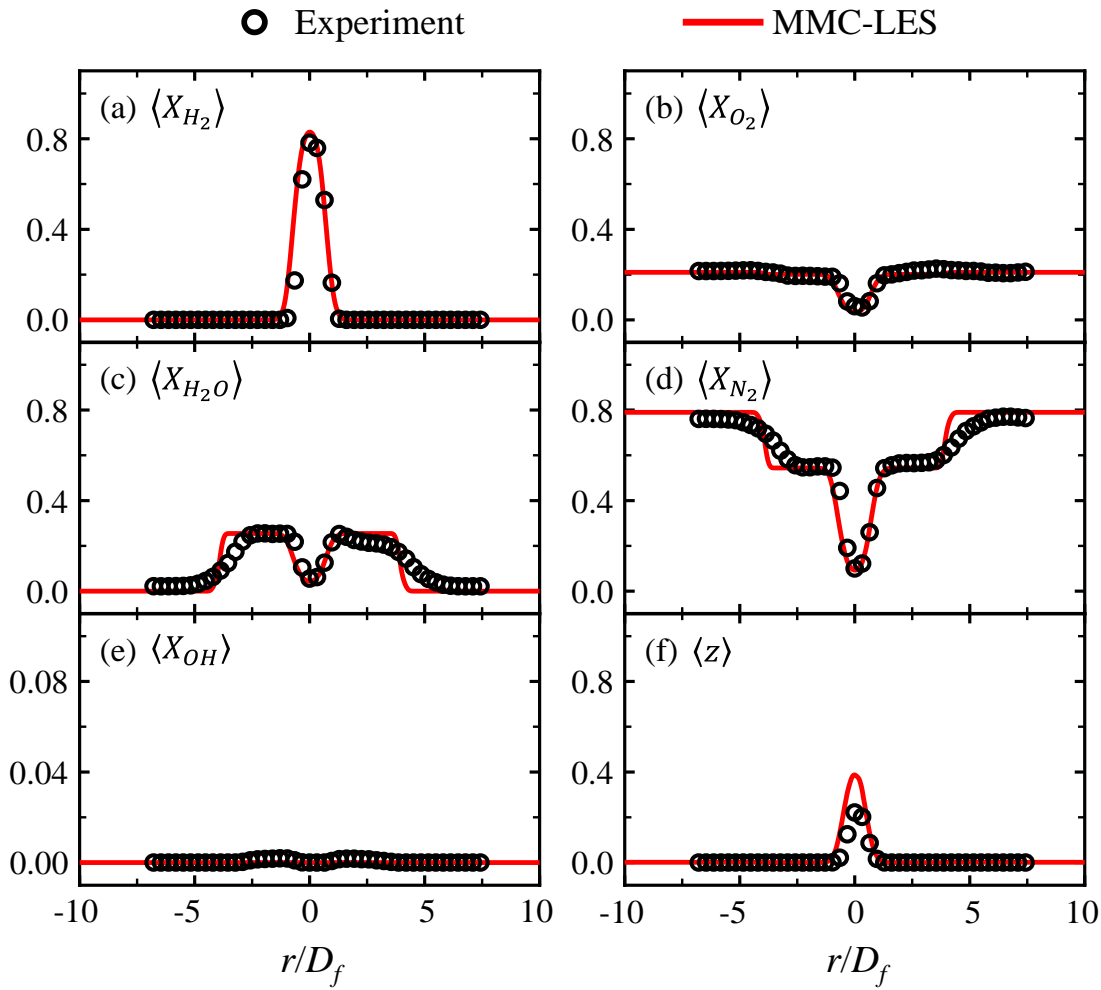

**Fig. 7.** Radial profiles of mean (a-e) species mole fractions and (f) mixture fraction at $x/D_f$ = 10.8. Experimental data from Ref. [31].

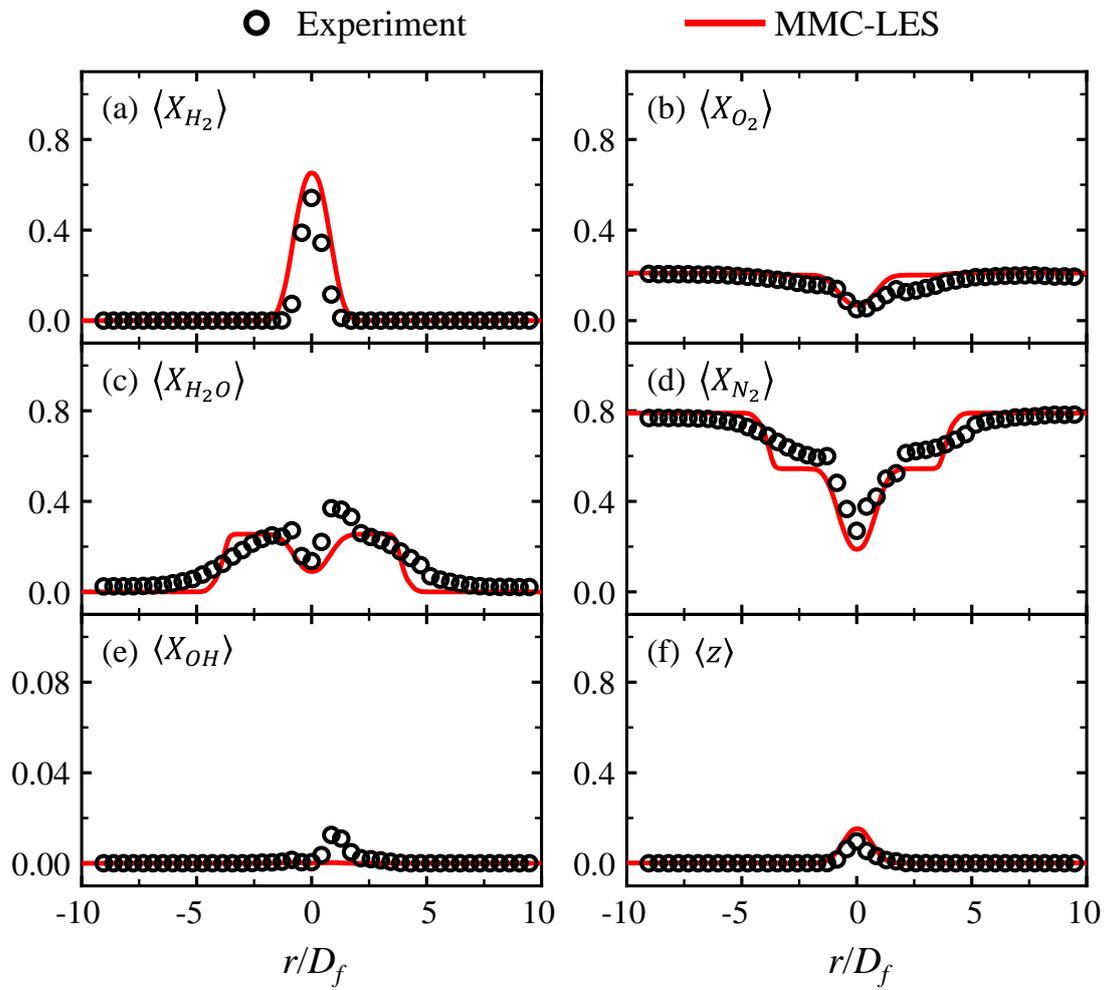

**Fig. 8.** Radial profiles of mean (a-e) species mole fractions and (f) mixture fraction at $x/D_f = 21.5$. Experimental data from Ref. [31].

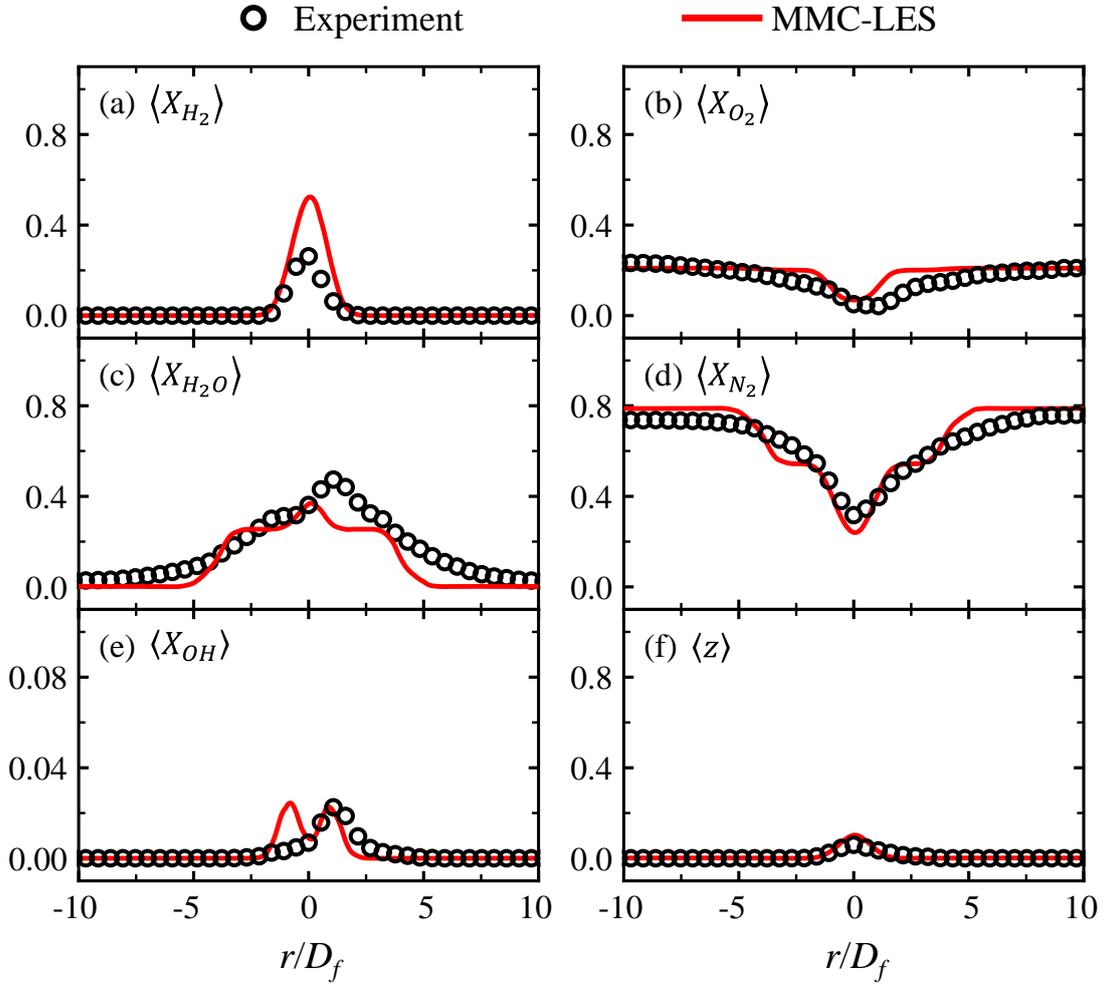

**Fig. 9.** Radial profiles of mean (a-e) species mole fractions and (f) mixture fraction at $x/D_f = 32.3$. Experimental data from Ref. [31].

### 5.3. Scalar statistics in mixture fraction space

Figures 10-12 show scatter plots of species mole fractions and temperature against the mixture fraction at three locations, $(x/D_f, r/D_f) = (10.8, 0.65), (32.3, 1.1)$ and $(43.1, 0)$, respectively. The mixing and chemically-equilibrium lines [31] are also shown. At $(x/D_f, r/D_f) = (10.8, 0.65)$ in Fig. 10, mixing dominates (below the appearance of autoigniting precursor $HO_2$, i.e., point 'a' at $x/D_f \approx 11$ in Fig. 3a) and therefore, the major species (i.e., $H_2$, $O_2$, $H_2O$ and $N_2$) closely follow their corresponding mixing lines. Furthermore, low OH (e.g., $X_{OH} < 10^{-4}$) may occur at both fuel-lean and fuel-rich conditions ($z = 0$-$0.05$). Similar observations are also found by Moule et al. [5] ($z = 0$-$0.06$), Zhang et al. [40] ($z = 0$-$0.055$) and Boivin et al. [54] ($z = 0$-$0.05$).

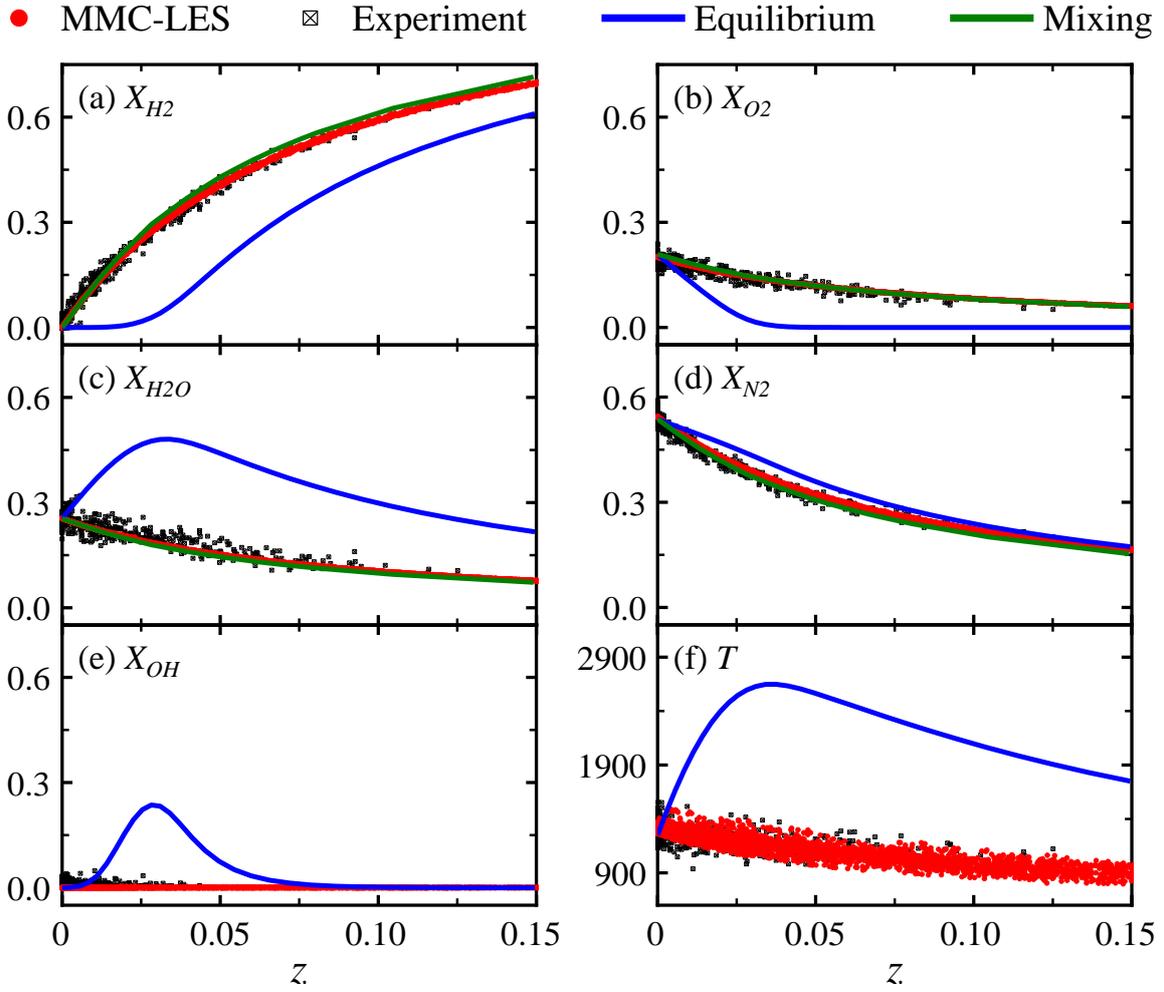

**Fig. 10.** Scatter plots of species mole fractions and temperature against mixture fraction at ($x/D_f$, $r/D_f$) = (10.8, 0.65). Experimental data from Ref. [31].

At ($x/D_f$, $r/D_f$) = (32.3, 1.1) in Fig. 11, the overall thermo-chemical state is reasonably captured by MMC-LES. The highest temperature is about 2,130 K in the reaction zone, below the equilibrium line. For OH radical, most stochastic data points produced by the MMC-LES are above the equilibrium line, except in the near-stoichiometric point. This may be caused by the strong turbulent mixing. These observations are also seen in Refs. [5,54] with a partially stirred reactor model and quasi-laminar chemistry method, respectively. Furthermore, the highest mixture fraction is about 0.15 in the simulation, higher than the experimental value of about 0.08. This is likely to have been caused by the insufficient radial spreading of the coflow into the ambient air, which leads to under-predicted dissipation of the fuel-containing coflow. However, when $z \gtrsim 0.06$ the scatter data of $X_{H_2}$, $X_{O_2}$, and $X_{H_2O}$ become more dispersive and move closer to their mixing lines. Our result is close to that of Moule et al. (the highest $z$ is about 0.13) [5], but is higher than that of Boivin et al. (the highest $z$ is about 0.07) [54].

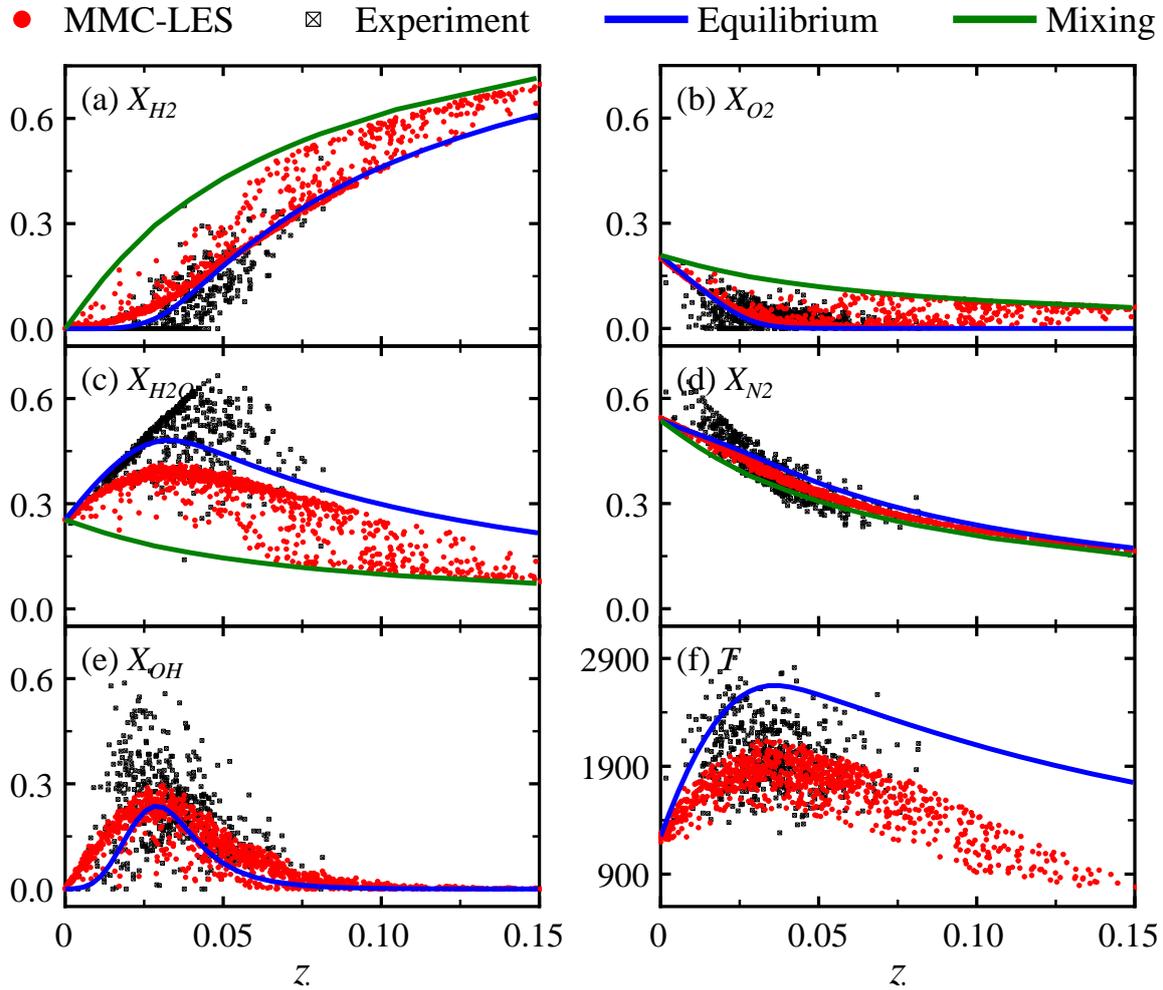

**Fig. 11.** Scatter plots of species mole fractions and temperature against mixture fraction at ($x/D_f$, $r/D_f$) = (32.3, 1.1). Experimental data from Ref. [31].

At ($x/D_f$, $r/D_f$) = (43.1, 0) in Fig. 12, the overall distributions of all scalar scatters are qualitatively similar to their counterparts at ($x/D_f$, $r/D_f$) = (32.3, 1.1) in Fig. 11, but move a bit closer to their equilibrium lines and become more dispersive because of additional turbulence development with downstream distance. This location lies in the turbulent combustion zone (IV in Fig. 2). Examples of sub-equilibrium temperature (i.e., $T$ in the flame zone is about 290 K lower than the equilibrium value) and super-equilibrium OH (i.e., $X_{OH}$ in the flame zone is higher than the equilibrium value) are also found in the experimental data. According to the Damköhler number analysis in the experimental work [31], this may be caused by the slow three-body recombination reactions.

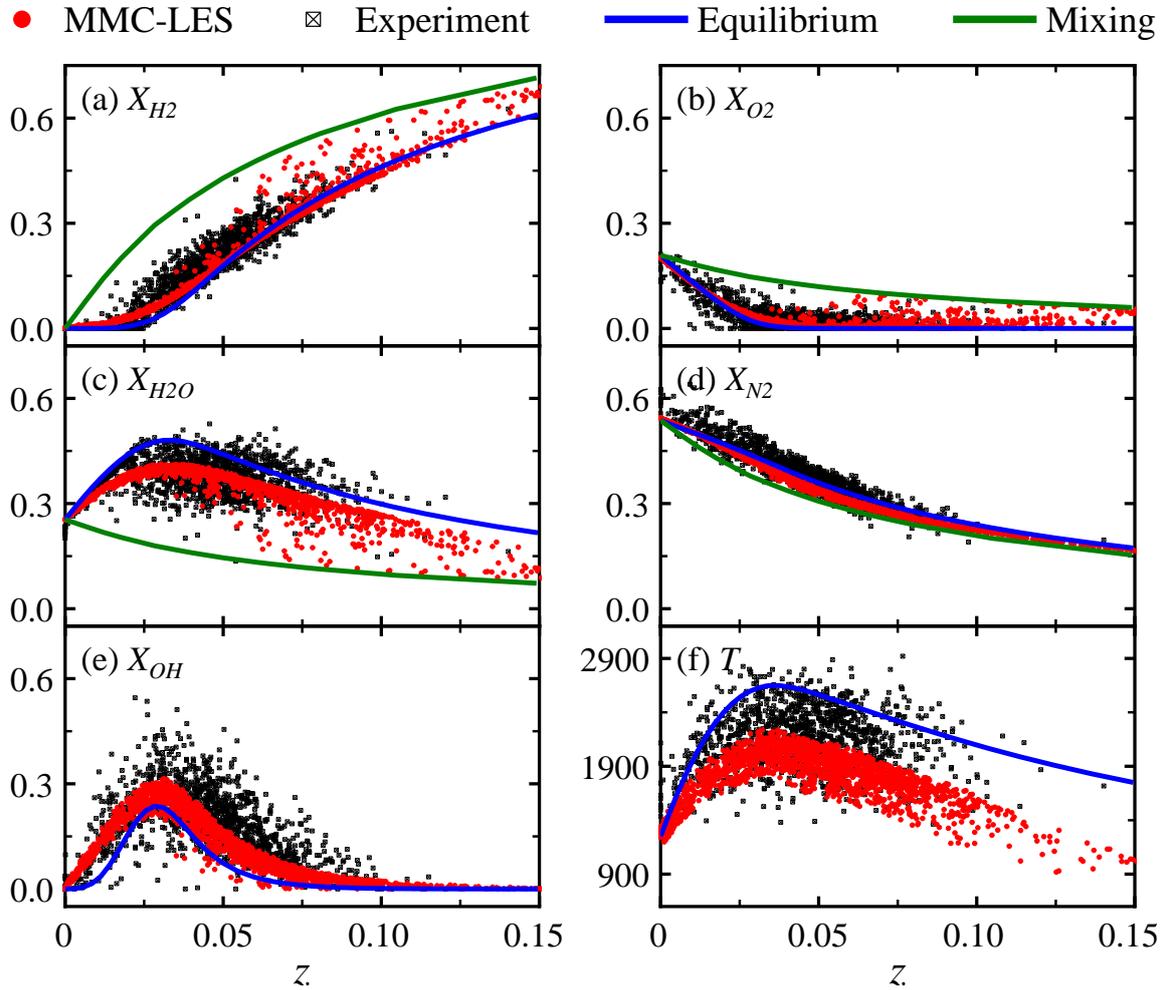

**Fig. 12.** Scatter plots of species mole fractions and temperature against mixture fraction at ($x/D_f$, $r/D_f$) = (43.1, 0). Experimental data from Ref. [31].

*5.4. Compressibility effects*

Incorporation of the pressure work and viscous heating effects is a significant model improvement for MMC in high-speed reacting flow simulations. To assess the modelling of pressure work and viscous heating, Fig. 13 shows the instantaneous distributions of the time derivative of standardised enthalpy ($\frac{dh^q}{dt}$), mixing term ($S_h^q$), pressure work ($W_P^q$) and the viscous heating term ($W_{VH}^q$) on the Lagrangian particles. The spatial distribution of $\frac{dh^q}{dt}$ in Fig. 13(a) is qualitatively similar to that of $S_h^q$ in Fig. 13(b) in most regions, suggesting that conditional subfilter mixing plays a dominate role in the variations of particle standardised enthalpy. As expected, in Fig. 13(c) the pressure work is important only near the shock / expansive waves. Also, in Fig. 13(d) it is evident that viscous heating is high (e.g., comparable to 10% of pressure work) mainly around the shear layers with large velocity gradient, e.g., the area labelled by 'G'. However, $W_{VH}^q$ is generally smaller than $W_P^q$ and $\frac{dh^q}{dt}$ by more than two orders of magnitude. This is also seen in our previous study for a model supersonic combustor [29], which is probably because of the low mixture viscosity (of the order of $10^{-5}$ Pa·s) but strong pressure derivative (as high as $10^{11}$ Pa/s or above) in these supersonic, shock-laden flows.

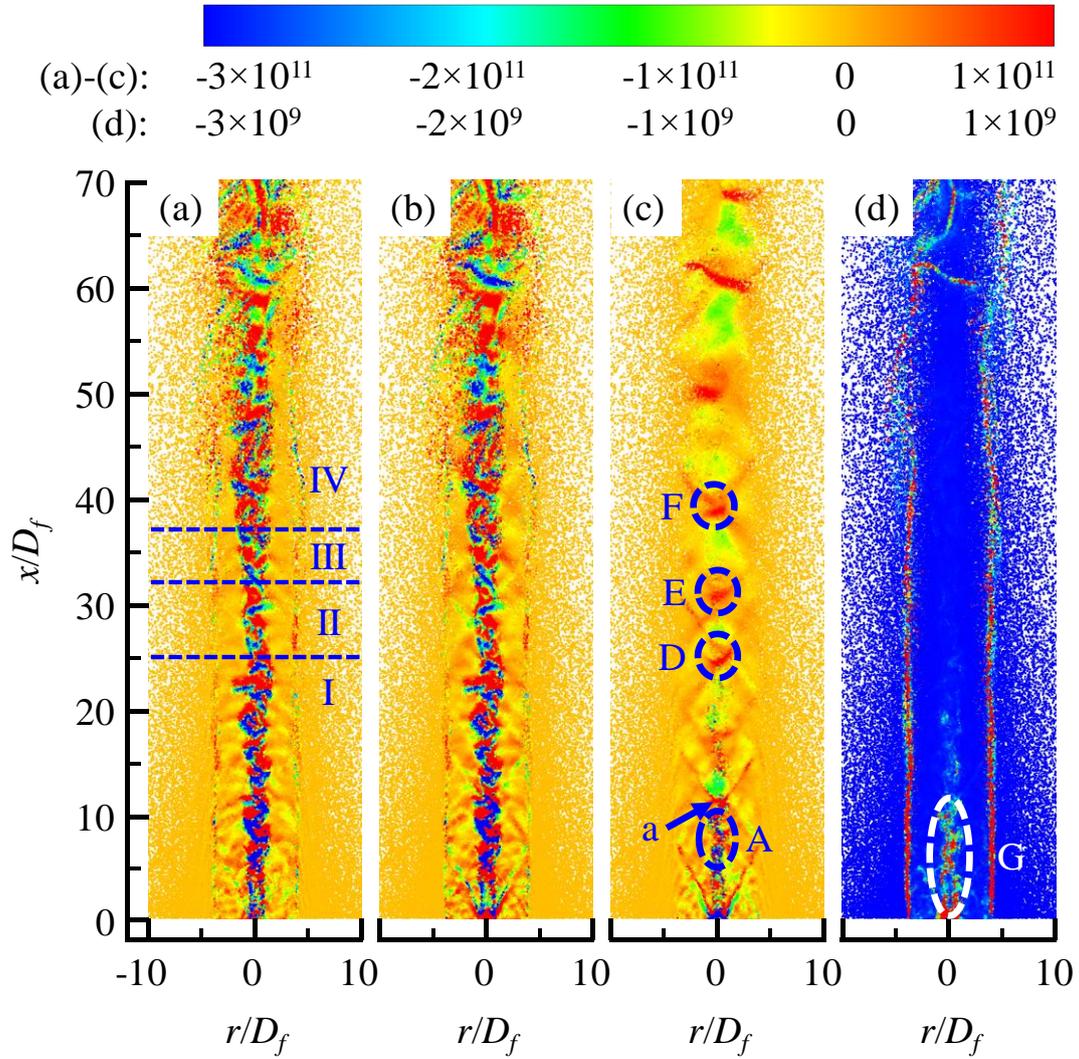

**Fig. 13.** Lagrangian particles colored by: (a) $\frac{dh^q}{dt}$, (b) $S_h^q$, (c) $W_P^q$ and (d) $W_{VH}^q$. All variables in W/kg. Zones I-IV, point a and ellipse A are identical to those in Fig. 3.

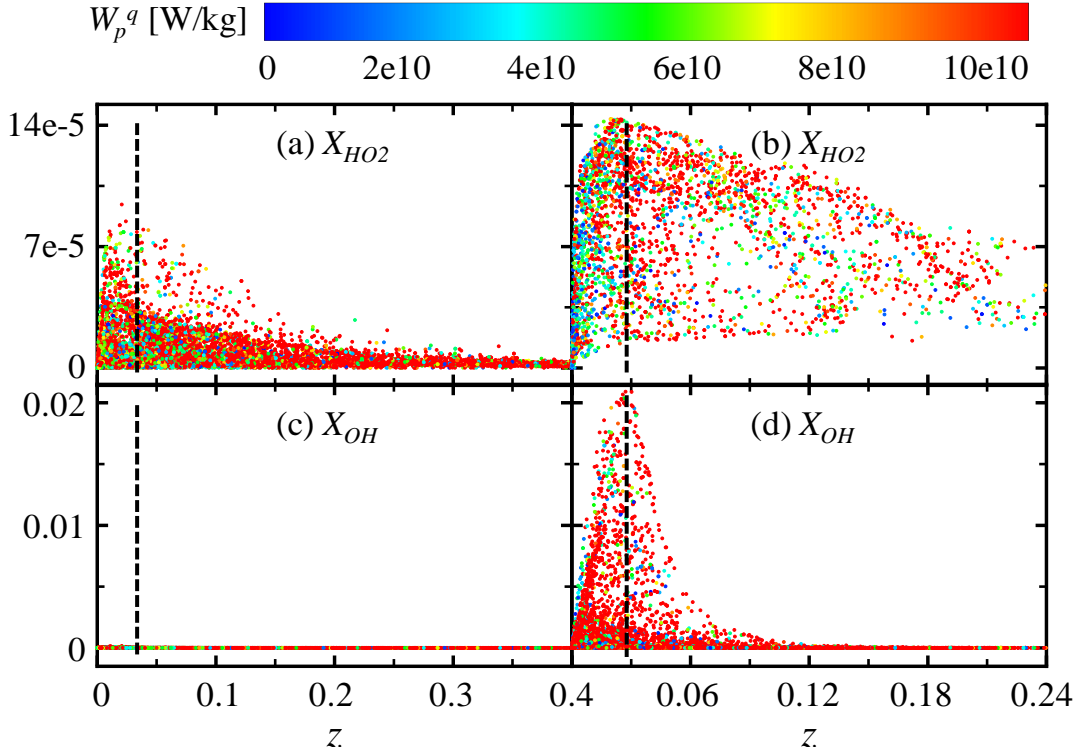

**Fig. 14.** Scatter plots of (a)-(b) $X_{HO2}$ and (c)-(d) $X_{OH}$ colored by pressure work. Results are from the Lagrangian particles in: $|r|/D_f \leq 4$, $9.5 \leq x/D_f \leq 11.5$ (first column); $|r|/D_f \leq 4$, $24.5 \leq x/D_f \leq 26.5$ (second column). Dashed lines: $z_{st} = 0.0297$.

The correspondence between chemical reaction and shock compression can be further confirmed in Fig. 14, which shows the scatter plots of $HO_2$ and $OH$ mole fractions against mixture fraction on the Lagrangian particles in two cylindrical domains ($|r|/D_f \leq 4$) in zones I and II, respectively. The scatter data are colored by pressure work. In Fig. 14(a), high $HO_2$ radical mole fraction (e.g., $X_{HO2} \approx 10^{-4}$) is observed near the stoichiometric line ($z_{st} = 0.0297$) and fuel-lean areas. This is related to the large $W_P^q$, which considerably promotes the autoignition. However, OH radical is limited (no flame occurs), as seen from Fig. 14(c).

In downstream locations near the flame stabilization point in Fig. 14(b), the peak $X_{HO2}$ is considerably increased. Furthermore, high $X_{HO2}$ is observed over a wider range of mixture fraction because of the extended ignition region (see Fig. 3a). Large pressure work can be seen for most data points (color scale), indicating the importance of pressure work for $HO_2$ generation in this area. In Fig.

14(d), compared to the results in Fig. 14(c), noticeable amounts of $X_{OH}$ radicals (e.g., $X_{OH} \approx 0.02$) are observed near the stoichiometric line, indicating the occurrence of flame there. The highest $X_{OH}$ generally occurs on those particles of large $W_p^q$, indicating the significance of shock compression (results in large positive $W_p^q$, see Fig. 13c) for the stabilization of the flame base at $x/D_f = 25$.

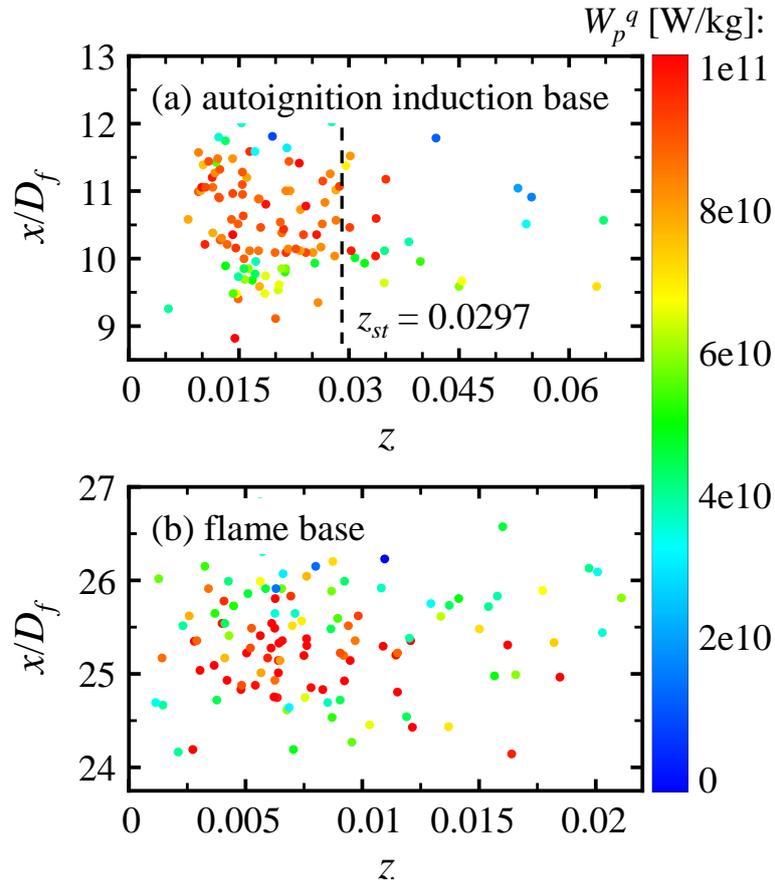

**Fig. 15.** Scatter plots of the (a) autoignition induction base and (b) flame base. Both are colored by pressure work.

Figure 15 further shows the evolutions of the instantaneous axial locations of Autoignition Induction Base (AIB) and Flame Base (FB). They are identified from the first occurrence of $X_{HO2} \geq 10^{-4}$ and $X_{OH} \geq 8 \times 10^{-3}$, respectively. Finite variations of these threshold values give almost the same results. The data are collected from 120 instants over 0.24 ms. It is seen that the AIB oscillates between $8.5D_f$ and $12.5D_f$, and those with high pressure work lie at $10D_f$ - $11.5D_f$ where the first shock intersection point a resides (see Fig. 3). There are also some scatter points from downstream locations with low $W_P^q$, which may result from hot coflow effects. Moreover, the AIB is mainly observed for lean mixtures with $z < z_{st}$. We can also find from Fig. 15(b) that the FB location varies at $24D_f$ - $27D_f$, implying that it fluctuates around the second shock intersection point 'b' (see Fig. 2b). Moreover, it is mainly located in the region $z \approx 0.005$-$0.01$, well below the stoichiometric mixture fraction. This is because the most reactive range of mixture fraction in supersonic combustion is considerably extended towards fuel-lean compositions due to the elevated pressures by shock compression [50,55,57]. Furthermore, $W_P^q$ is significant around $x/D_f \approx 24.8$-$26$, which means that the pressure work also plays an important role for stabilization of the lifted supersonic hydrogen flame.

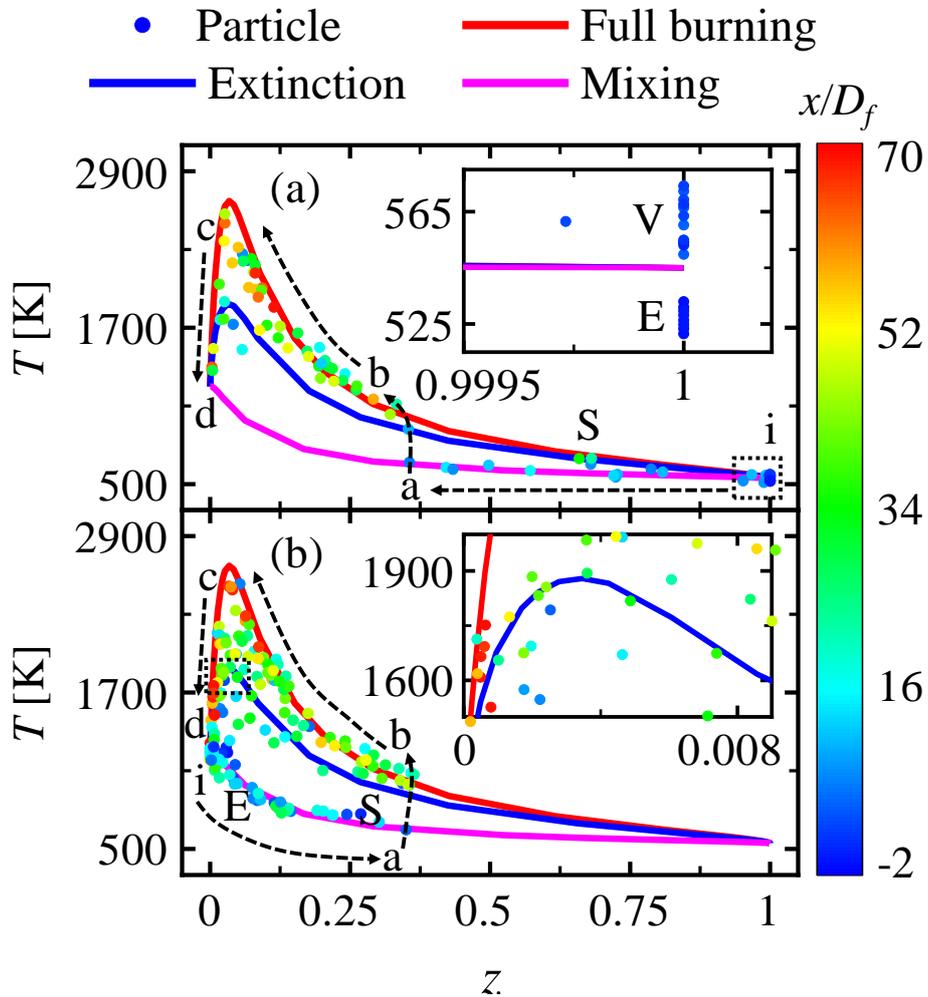

**Fig. 16.** Trajectories of particles (a) P1 and (b) P2 in temperature−mixture fraction space colored by location. Lines represent the full burning (red), extinction (blue) and mixing (pink) solutions. Arrows: particle evolution direction.

*5.5. Lagrangian particle trajectory*

In this section, trajectories of stochastic Lagrangian particles from the MMC-LES are extracted to investigate the reactive scalar evolutions subject to the supersonic flow fields. Four representative stochastic particles are tracked, which are respectively injected from the fuel jet ($r_0 \approx 0.005$ mm), fuel jet shear layer ($r_0 \approx 1.2$ mm), and the coflow ($r_0 \approx 2.0$ and 6.0 mm). They are also marked in Fig. 1(b). Hereafter, they are respectively termed as P1, P2, P3 and P4.

Figure 16 shows the temperature evolutions of particles P1 and P2, colored by their instantaneous streamwise positions. Laminar flame temperatures calculated using OPPDIF [58] under three different strain rates are also shown for reference. The line with strain rate, $S_r \approx 19$ /s is deemed the full burning state, whereas the line with $S_r \approx 3,971$ /s is the "extinction line", below which the particles are extinguished. The mixing line is also shown. Note that in *T-z* space, reactions can only move the particles vertically (i.e., change the particle temperature) because mixture fraction is conserved during chemical reactions. However, mixing makes the particles move both vertically and horizontally, corresponding to the variations of temperature and mixture fraction of the particles. In contrast to low-speed flows [59], in supersonic flows the particle temperature variations may also be associated with shocks / expansion waves, and viscous heating. Hence, they can also move the particle positions vertically in *T-z* space.

In Fig. 16(a), particle P1 (starts from $z = 1$) is first affected by viscous heating in the nozzle and shock compression near the nozzle exit. Its temperature is increased to 574 K ('V' in the inset). Then the particle experiences expansion and its temperature is decreased to 522 K ('E'). The mixture fraction decreases because of mixing with oxidizer particles from points 'I' to 'a' and *T* evolves nearly along the mixing line. However, *T* may also be raised above the mixing line (marked by 'S') due to shock and/or viscous heating. From points 'a' to 'b', autoignition occurs at $z \approx 0.354$ (i.e., fuel-rich) and *T* rises nearly vertically along line a-b because of chemical reaction at about $x/D_f = 6.8$. From points 'b' to 'c', *T* changes roughly following the full burning line. However, interactions with the shocks, expansion waves and viscous heating result in some variations. These effects are not observable in subsonic flames [59], and in turn, make the particle trajectory deviate from the full burning or mixing lines in supersonic flames. From points 'c' to 'd', the fully burned particle ($z \approx 0$) mixes with surrounding oxidizer particles and $T$ decreases towards 1,250 K. Overall, the mixture fraction of the fuel particle can vary in its life history from 0 to 1. According to Wang et al. [59] and Mitarai et al. [60], this particle trajectory is regarded as continuous burning, most of which lies within the burning region (above the extinction line) once it is ignited.

In Fig. 16(b), particle P2 (starts from $z = 0$) originates in the recirculation zone between the fuel and coflow nozzle walls (see Fig. 1b). Therefore, it has relatively long time to mix with the fuel particles before it is ignited, and viscous heating on this particle is weak because of the small velocity gradient. From points 'i' to 'a', the particle proceeds with expansion and shock compression ('E' and 'S') when it mixes with the fuel particles near the mixing line. At point 'a' ($z \approx 0.35$ and $x/D_f \approx 7.3$), it is ignited, and $T$ increases rapidly along line a−b. Then it burns near the full burning line until point 'c'. After that, it mixes with oxidizer particles and $T$ decreases. The burning process (lines b-c) of P2 are different from P1 because the former intermittently enters the extinction state (see the inset). The mixture fraction of particle P2 varies in the range of 0 to 0.35.

Figure 17 shows the temperature evolutions of particles P3 and P4. In Fig. 17(a), particle P3 from the coflow is first heated to 1,430 K ('V' in the inset) because of shock compression and viscous heating. Then $T$ decreases to 1,058 K ('E') because of expansion after it leaves the nozzle. From points 'i' to 'a', $T$ can be raised above the mixing line by shock compression (e.g., 'S') although only mixing proceeds. At point 'a' ($z \approx 0.257$ and $x/D_f \approx 9.8$), the particle is ignited and $T$ increases rapidly along line a-b. From points 'b' to 'c', the particle burns around the burning line under the effects of shocks, expansion waves or viscous heating. From points 'c' to 'd', $T$ is decreased because of the mixing with oxidizer particles ($z \approx 0$). This particle experiences localized extinctions in the burning process along line b-c. The mixture fraction varies in the range of 0 to 0.257.

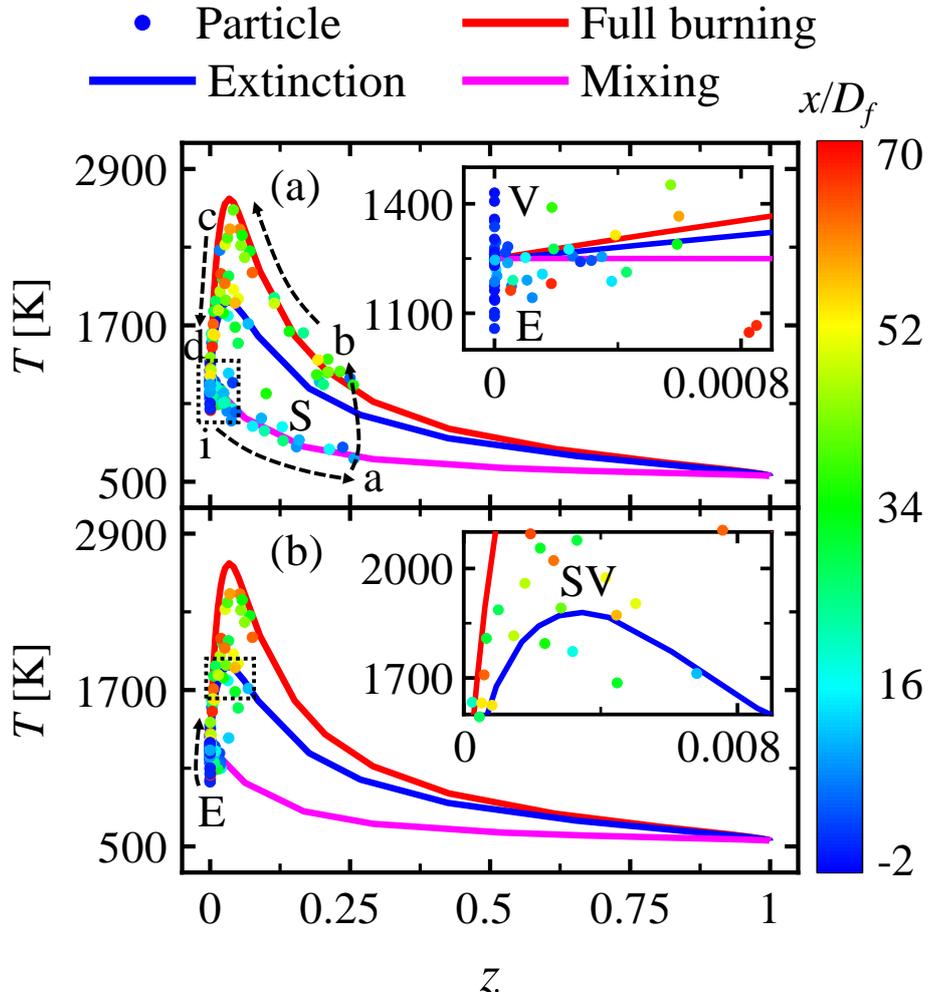

**Fig. 17.** Trajectories of particles (a) P3 and (b) P4 in temperature−mixture fraction space. Arrows: particle evolution direction.

In Fig. 17(b), particle P4 (from $z = 0$) from the outer part of coflow first experiences expansion (region 'E') after it is injected in the domain. Viscous heating in the nozzle is not significant as it is relatively far away from the nozzle wall. Note that this particle is difficult, if not impossible, to directly mix with a fuel particle. Therefore, its trajectory generally deviates far from the mixing line. However, when it travels downstream in the physical space, it still can mix with fully burned (of the highest temperature but nearly zero mixture fraction) or partially burned (of intermediate temperature but is still fuel-containing) particles. In this manner, both its temperature and mixture fraction can be increased. Furthermore, under the effects of shock compression, viscous heating and mixing with high temperature particles, it may enter the burning region (zone 'SV'). Overall, extinction dominates for most segment of its trajectory and the mixture fraction ranges in 0 to 0.071 (fuel-lean).

*5.6. Chemical explosive mode and timescale*

The chemical explosive mode analysis [32] is performed on the Lagrangian particles in MMC. Key combustion features can be identified through eigen-analyses of the local chemical Jacobian matrix, $J_c$. Chemical modes are associated with the eigenvalues of $J_c$, among which the one with the maximum real part $\lambda_e$. A Chemical Explosive Mode (CEM) is identified when the real part of $\lambda_e$, $Re(\lambda_e)$, is positive. In this section, it is visualized through

$$\lambda_{cem} = max\{sign[Re(\lambda_e)], 0\} \cdot log_{10}[1 + |Re(\lambda_e)|], \tag{13}$$

where *sign(x)* is the sign function, whilst *max(x,y)* is the maximum function. More details of CEMA can be found in Ref. [32] and are not repeated here. Moreover, explosion index [61] is used to quantify the contributions of temperature and species concentrations to CEM in Lagrangian particles, i.e.,

$$EI_T^P = EI_T \cdot max\{sign[Re(\lambda_e)], 0\}, \tag{14}$$

$$EI_{Y_i}^P = EI_{Y_i} \cdot max\{sign[Re(\lambda_e)], 0\}, \tag{15}$$

where $EI_T$ and $EI_{Y_i}$ are respectively the explosion indices of temperature and species obtained from the CEMA (which may be non-zero in non-CEM regions).

Figure 18(a) shows the spatial distribution of chemical explosive mode $\lambda_{cem}$ of the Lagrangian particles. It is seen that CEM particles are present immediately after the burner exit, mainly in the shear layer between the fuel jet and hot coflow. When extensive $HO_2$ radicals occur after point a (see Fig. 3a), the explosive particles grow rapidly and also become significant in the central fuel jet until zone IV, above which they vanish in most areas but are only present near the fuel shear layer. To quantify the relative importance of temperature and species to particle CEM, the explosion indices for temperature, H and OH, as well as the iso-lines of instantaneous stoichiometric reference mixture fraction are shown in Figs. 18(b)-18(d). In Fig. 18(b), temperature contributes to CEM mainly inside the fuel jet, especially below zone IV, indicating the *thermal explosion* propensity in these areas. In Figs. 18(c) and 18(d), the H and OH radicals are dominant in CEM (therefore *radical explosion* propensity) respectively in the fuel-rich and fuel-lean sides near the fuel jet shear layer. This can be justified through the competition of elementary reactions. On the fuel-rich side of the fuel / coflow shear layer, the hydrogen dissociation reactions are dominant (e.g., $H_2 + M \rightarrow H + H + M$ in the hydrogen mechanism [52]). However, on the fuel-lean side, the reactions related to oxidizer dissociation are important (e.g., $OH + OH \rightarrow O + H_2O$). Therefore, H and OH radicals play dominant roles in CEM respectively in the fuel-rich and fuel-lean regions. Similar qualitative distributions of the stratified explosion indices are also observed for a near-wall transverse hydrogen flame [32] and a lifted ethylene jet flame in a heated air coflow [62].

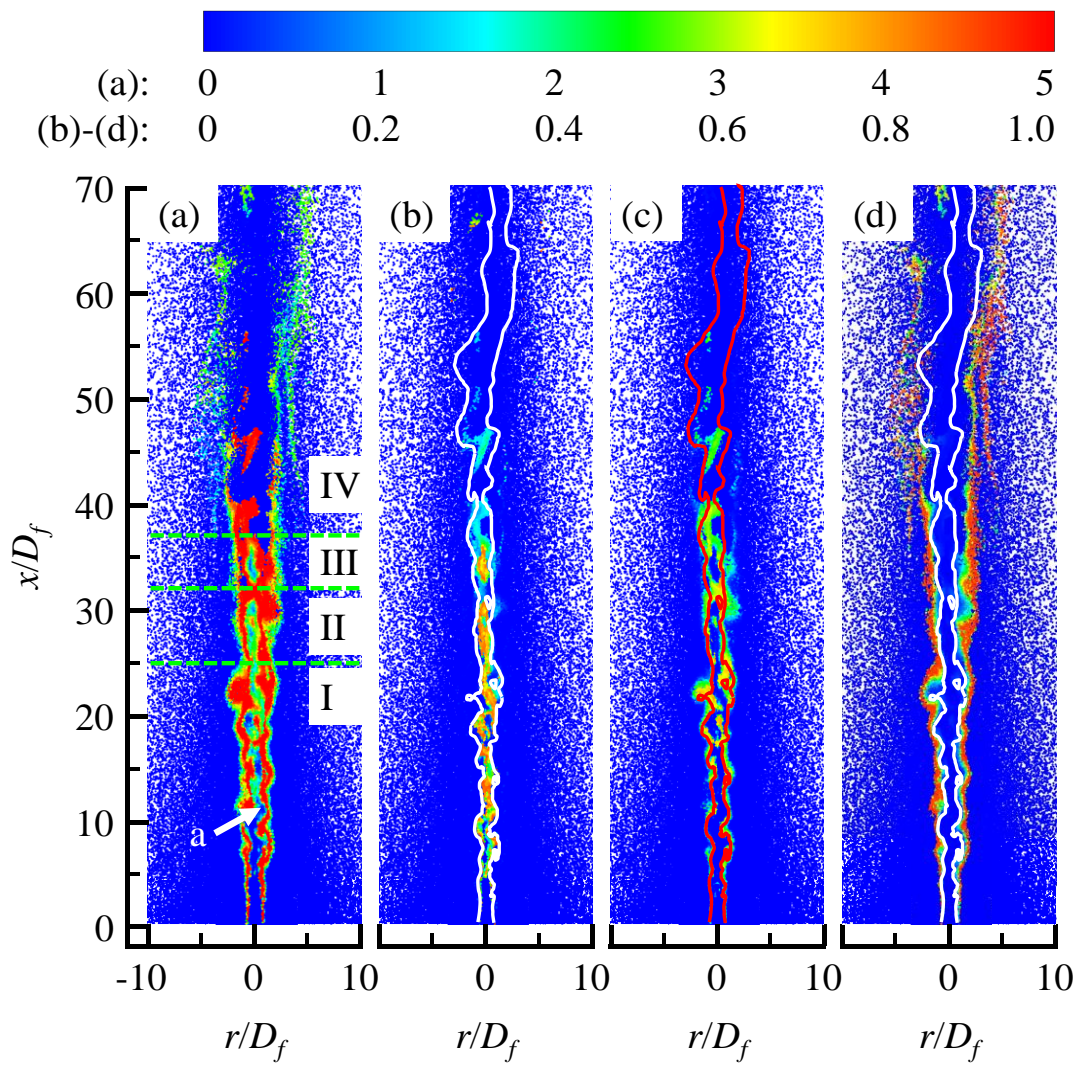

**Fig. 18.** Lagrangian particles colored by: (a) $\lambda_{cem}$, (b) $EI_T^P$, (c) $EI_{Y_H}^P$ and (d) $EI_{Y_{OH}}^P$. Iso-lines in (b)-(d) represent instantaneous stoichiometric mixture fraction.

The CEMA also provides the accurate chemical timescale information and thereby Lagrangian particle Damköhler number can be derived from [61]

$$Da = \tau_L/\tau_c = \tau_L \cdot \lambda_e, \quad (16)$$

where $\tau_c = 1/\lambda_e$ is the chemical timescale from the CEMA, and $\tau_L$ is a mixing timescale estimated from the a-ISO model [38] in MMC, i.e.,

$$\tau_L = \beta \cdot \frac{C_f}{C_L} \cdot \frac{d_{\tilde{f}}^2}{\chi}, \quad (17)$$

where $C_f = 0.1$, $C_L = 1$ and $\beta = 3$ are model constants [22]. $d_{\tilde{f}}$ is the particle mixing distance in $\tilde{f}$-space, and $\chi$ is the scalar dissipation rate. Under the assumption that both LES mesh and particle mixing distance in $x$-space are within the inertial range, $\chi$ is universal and calculated as [22]

$$\chi = 2\mathcal{D}_{eff}\left(\nabla \tilde{f}\right)^2, \quad (18)$$

where $\nabla \tilde{f}$ is the gradient of the RMF (solved on LES mesh in Eq. 5) interpolated to the particle location.

Figure 19 shows the instantaneous distributions of Lagrangian particle Damköhler number. Generally, large positive $Da$ indicates that the mixture is autoigniting, whereas negative $Da$ (in logarithmic scale) indicates that the mixture is non-ignitable or post-ignited [62]. In Fig. 19(a), $Da$ is well above unity (~100) in the fuel jet shear layer, right after the burner exit. This is mainly attributed to the small chemical timescale of the mixture (large $\lambda_{cem}$ in Fig. 18a). The mixing between the fuel jet and high-temperature coflow is important for the mixture to achieve high reactivity in the reaction induction period. This is more appreciable in Fig. 19(b), which shows $Da$ for $x/D_f$ = 9.5-11.5 (in zone I). One can see that $Da \approx 100$ in the shear layer, however it is much lower in most part of the fuel jet center. Only after $x/D_f \approx 10.5$, $Da > 1$ is observed intermittently along the centerline, mainly caused by local shock compression, which decreases the chemical timescale. In Fig. 19(c), $Da$ is also small around the centerline for $x/D_f$ = 23-25 (the end of zone I), near the flame stabilization point ($x/D_f$ = 25). However, the region with high $Da$ (e.g., $Da \geq 10$) is pronounced off the centerline. In Fig. 19(d), $Da$ becomes significant in the jet center for $x/D_f$ = 25-27 (zone II). The rapid increase of $Da$ in the jet center is caused by the shock intersection here (i.e., point b in Fig. 2b). In Fig. 19(e), part of the central jet and fuel shear layer reaches the post-ignition state (i.e., those of $Da \ll 1$) for $x/D_f$ = 33-35 (zone

III). In Fig. 19(f), extensive zones with $Da \ll 1$ occur in the jet center as well as in the fuel-lean side of the fuel jet shear layer for $x/D_f = 38$-$40$ (zone IV). Further downstream, $Da$ decreases rapidly because the mixture is no longer chemically explosive (relatively large chemical timescale), as seen from Fig. 18(a).

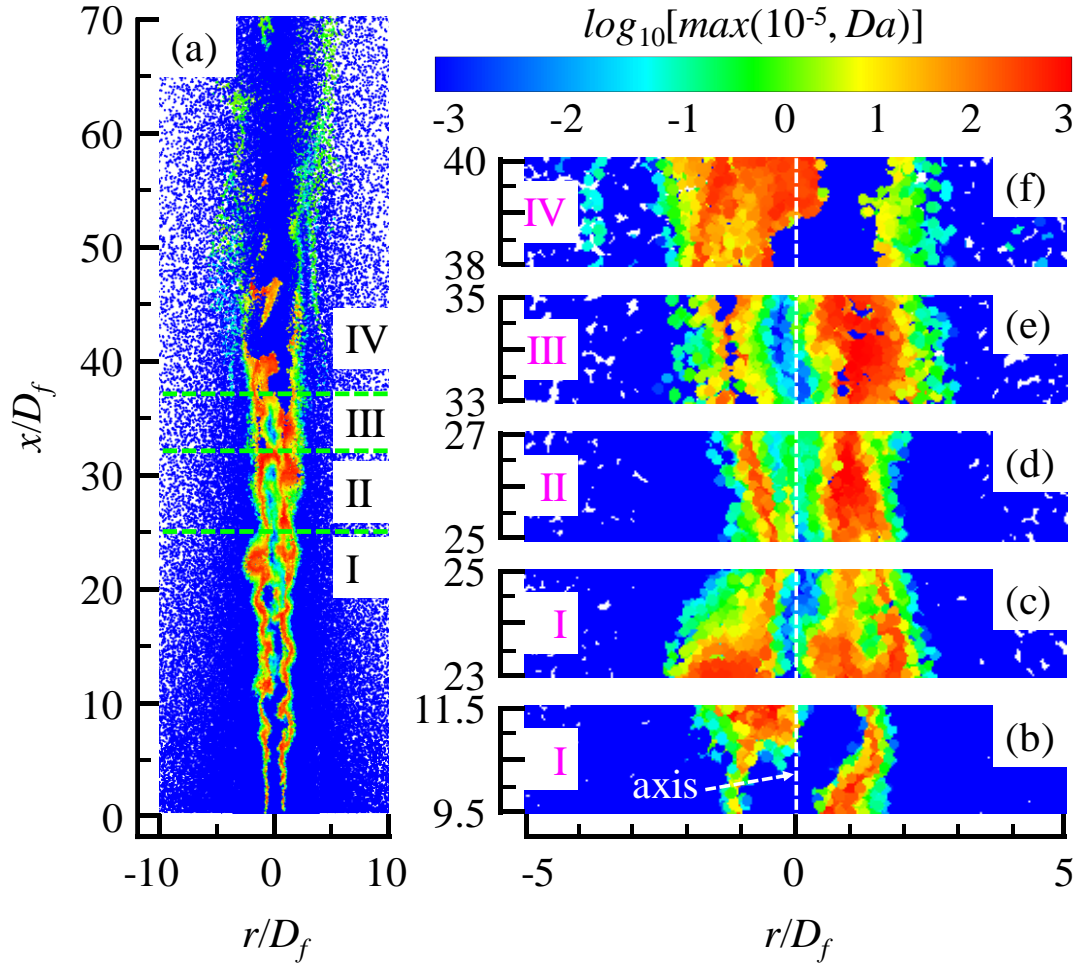

**Fig. 19.** Lagrangian particles colored by Damköhler number (in logarithmic scale). (a) is on the center plane, (b)-(f) are from subdomains of $x/D_f = 9.5$-$11.5$, $23$-$25$, $25$-$27$, $33$-$35$ and $38$-$40$ with $|r|/D_f \leq 5$, respectively.

Figure 20 further shows the axial profiles of averaged Lagrangian particle Damköhler numbers, which are estimated based on the particles within a cylindrical domain ($|r|/D_f \leq 0.01$) along the jet centerline. Those estimated by the experimentalists at different locations (some of which are off the centerline) [31] are also shown for reference. In MMC-LES, upstream $Da$ is slightly above zero in the jet center, which increases slowly until $x/D_f \approx 11$. After that, $Da$ increases rapidly because of shock intersection and sufficient mixing with the hot coflow. It peaks at the flame stabilization point ($x/D_f = 25$) and sustains high level in zones II-III. After zone III, it decreases rapidly because that central fuel jet is almost fully burned. However, discrepancies exist for $Da$ beyond $x/D_f = 30$, which can be attributed to the constant chemical timescales used in their estimations [31]. The streamwise evolution of $Da$ is similar to the DNS work of subsonic lifted hydrogen [61] and ethylene [62] flames.

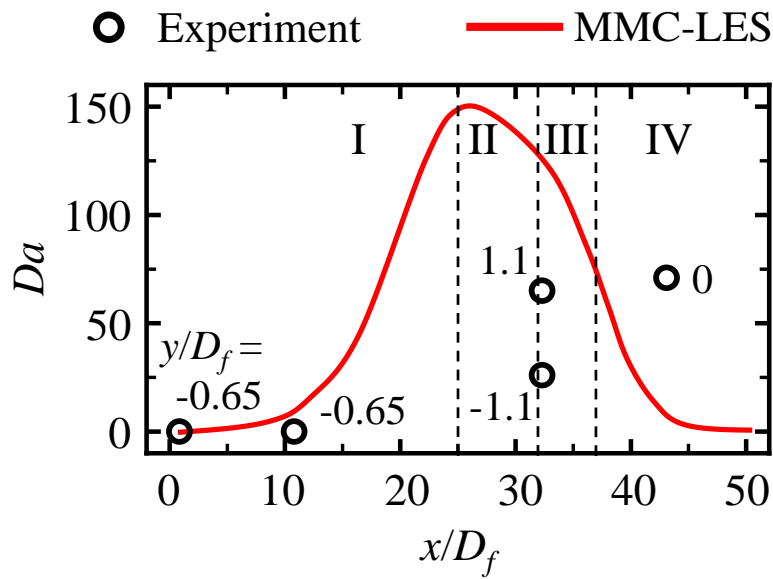

**Fig. 20.** Axial profile of particle Damköhler number along $r/D_f = 0$. Experimental data from Ref. [31]. I-IV: flame zones in Fig. 2(c).

## 6. Conclusions

A lifted supersonic hydrogen jet flame in vitiated coflow is simulated with sparse MMC-LES model for fully compressible flows. The pressure work and viscous heating terms are incorporated in the MMC, to model the interactions between supersonic flow structures (e.g., shock waves) and chemical reactions.

The results show that the diamond-shaped alternative expansion / shock wave structures, shock-flame interaction, and overall flame characteristics (e.g., lift-off, staged flame evolution) are accurately reproduced. Moreover, the MMC-LES results show reasonable agreement with the experimental data in terms of time-averaged quantities (axial velocity, temperature, species mole fractions and mixture fraction) and the available root-mean-square values. Furthermore, the scatter plots of species mole fractions and temperature against mixture fraction also agree satisfactorily with the experimental counterparts.

The effect of pressure work and viscous heating on autoignition initiation and supersonic flame stabilization is also assessed. The positive contribution of pressure work to the variation of particle standardised enthalpy at the shock intersection point enhances the buildup of $HO_2$ radical in the reaction induction and autoignition initiation stages. Furthermore, pressure work contributes positively to the production of OH radical in the transitional and turbulent combustion zones. Nonetheless, viscous heating is negligibly small in most regions expect those with strong velocity gradients (e.g., around the burner exit), compared to particle enthalpy variation and pressure work. The autoignition induction base and flame base oscillate around two shock intersection points, considerably enhanced by high pressure work.

The evolutions of particle information, e.g., temperature and mixture fraction, are extracted through particle trajectory analysis to discuss the interactions between particle and gas dynamics. For a particle from the central jet, it first mixes with the oxidizer particles under the effects of shock, expansion wave and viscous heating. Then it is autoignited and almost continuously reactive along the full burning line until the fuel is consumed. For representative oxidizer particles, they first mix with other fuel-containing particles and are autoignited because of shock compression and/or hot particle heating after some distance of reaction induction. However, during the burning process the particles may also intermittently enter the extinction region. The shock, expansion wave and/or viscous heating effects in supersonic flows can deviate the particle trajectories from the full burning or mixing lines, which are not observed in subsonic flames.

The chemical explosive mode analysis is performed based on the Lagrangian particles. The results show that temperature (may vary because of mixing, shock compression, flow expansion and viscous heating) contributes dominantly to CEM in the central fuel jet. The H and OH radicals are respectively dominant in the fuel-rich and fuel-lean sides near the fuel jet shear layer. The Lagrangian particle

Damköhler numbers show that large positive *Da* first occurs in the fuel jet / coflow shear layer, right after the burner exit. Then *Da* is enhanced at the first shock intersection point in the jet center and peaks around the flame stabilization point, right after the second shock intersection point.

## Acknowledgement

This work was supported by MOE Tier 1 grant (R-265-000-653-114) and the USyd-NUS Partnership Collaboration Award. The simulations are performed with the computational resources from National Supercomputing Center Singapore (NSCC, https://www.nscc.sg/). ZR is supported by National Natural Science Foundation of China 52025062. Wantong Wu from Tsinghua University is thanked for assistance with the chemical explosive mode analysis. Professors T.-S. Cheng and R.W. Pitz are acknowledged for sharing the experimental data and burner details.